\documentclass[english,12pt]{article}

\usepackage[left=1in,right=1in,bottom=1in,top=1in]{geometry}

\usepackage[utf8]{inputenc}
\usepackage{babel}
\usepackage{slantsc}
\usepackage{array}
\usepackage{amsmath}
\usepackage{amsthm}
\usepackage{amssymb}
\usepackage{graphicx}
\usepackage{enumerate}
\usepackage{mathtools}
\usepackage{natbib}
\usepackage{bbm}
\usepackage{tikz}
\usepackage{multirow}
\usetikzlibrary{tikzmark}
\usepackage{subfigure}
\usepackage{xspace}
\usepackage{bigints}
\usepackage{booktabs}
\usepackage{thmtools}
\usepackage{setspace}
\usepackage{fancybox,framed}
\usepackage{multicol}

\usepackage{pgf}
\usepackage{pgfplots}
\pgfplotsset{compat=newest} 
\pgfplotsset{plot coordinates/math parser=false} 
\newlength\figureheight 
\newlength\figurewidth 

\usepackage{heuristica}
\usepackage[heuristica,vvarbb,bigdelims]{newtxmath}
\usepackage[T1]{fontenc}
\usepackage{textcomp} 
\usepackage{fix-cm}

\newtheorem{theorem}{Theorem}
\newtheorem{proposition}{Proposition}

\newtheorem{corollary}{Corollary}

\newtheorem{lemma}{Lemma}

\newtheorem*{theorem*}{Theorem}
\theoremstyle{definition}
\newtheorem{definition}{Definition}
\declaretheorem[style=definition]{example}
\renewcommand\thmcontinues[1]{continued}

\usepackage[labelsep=period]{caption}

\DeclareMathOperator*{\supp}{supp}

\makeatletter
\newcommand{\subalign}[1]{%
  \vcenter{%
    \Let@ \restore@math@cr \default@tag
    \baselineskip\fontdimen10 \scriptfont\tw@
    \advance\baselineskip\fontdimen12 \scriptfont\tw@
    \lineskip\thr@@\fontdimen8 \scriptfont\thr@@
    \lineskiplimit\lineskip
    \ialign{\hfil$\m@th\scriptstyle##$&$\m@th\scriptstyle{}##$\crcr
      #1\crcr
    }%
  }
}
\makeatother

\definecolor{PennBlue}{RGB}{001,031,091}
\definecolor{PennRed}{HTML}{6e2f35}
\usepackage{hyperref}
\hypersetup{
	pdfborder = {0 0 0},
    colorlinks,
    citecolor=PennRed,
    filecolor=PennRed,
    linkcolor=PennRed,
    urlcolor=PennRed,
    breaklinks=true
}
\definecolor{CobaltBlue}{HTML}{3AAFA9}
\definecolor{Raspberry}{HTML}{cb3d58} 
\definecolor{BrightEmerald}{HTML}{00883A}
 
\usepackage[blocks]{authblk}   
\usepackage{datetime2}         
\title{\LARGE The Empirical Content of Bayesian Updating\\ under Misspecification}      
\author{\normalsize Pooya Molavi}         
\affil{\small Northwestern University}     
\date{\small March 2026}

\vfuzz \hfuzz
\allowdisplaybreaks

\newcolumntype{C}[1]{>{\centering\arraybackslash$}p{#1}<{$}}
\newlength{\mycolwd}

\usepackage[protrusion=true,expansion=true,final,babel]{microtype}
\begin{document}

\maketitle

\begingroup
  \renewcommand\thefootnote{}      
  \footnotetext{\footnotesize Acknowledgments: I am grateful to Drew Fudenberg, Alexander Jakobsen, Peter Klibanoff, Alessandro Pavan, Alvaro Sandroni, Eran Shmaya, Alireza Tahbaz-Salehi, and participants at the 2025 ASSA meeting for their valuable comments. Earlier versions of this paper were distributed under the titles ``The Empirical Content of Bayesianism'' and ``Misspecified Bayesianism.''}
  \footnotetext{\footnotesize Email: \href{mailto:pmolavi@kellogg.northwestern.edu}
                           {pmolavi@kellogg.northwestern.edu}}
\endgroup

\setcounter{footnote}{0}          
\renewcommand{\thefootnote}{\arabic{footnote}}

\begin{abstract}
\fontsize{11}{11}\selectfont \baselineskip0.58cm
\noindent 
An agent is a misspecified Bayesian if she updates her belief using Bayes' rule given a subjective, possibly misspecified model of her signals. This paper shows that a belief sequence is consistent with misspecified Bayesian updating if and only if the set of posteriors admits a countable partition such that the prior \emph{contains a grain} of the conditional average posterior on each cell. The condition imposes essentially no restrictions on posteriors given a full-support prior over a finite state space and reduces to a support inclusion condition on compact state spaces under mild regularity assumptions. However, it rules out posterior beliefs with heavier tails than the prior on unbounded state spaces. In Gaussian environments, it implies that posterior uncertainty cannot exceed prior uncertainty. The results delineate the boundary between updating rules that are observationally equivalent to Bayesian updating under misspecification and genuinely non-Bayesian rules. As an application, the paper shows that diagnostic expectations are consistent with misspecified Bayesianism, whereas some parameterizations of smooth diagnostic expectations are not.
\end{abstract}

\thispagestyle{empty}
\newpage 
\setcounter{page}{1}\fontsize{12}{12}\selectfont\baselineskip0.69cm


\section{Introduction}
Following the treatise of \citet*{savage1972foundations}, the Bayesian theory of probability has become the dominant paradigm in the modeling of decision making under uncertainty. This paradigm's dominance in economics is not unwarranted. It allows one to assign probabilities to unique or rare events; it has an elegant foundation in the study of rational choice under uncertainty; and it is appealing from a normative perspective---as \citet*{epstein1993dynamically} proclaim, ``dynamically consistent beliefs must be Bayesian.''

Whether Bayesian updating is an accurate positive model of how individuals actually revise their views is a different matter. A large body of evidence documents violations of Bayesian updating (e.g., \citealp*{coibion2015information, bordalo2020overreaction}). However, standard tests of Bayes' rule (e.g., \citealp*{augenblick2021belief}) are joint tests of Bayesian updating and the assumption that agents have correctly specified models of the data-generating process. Therefore, any rejection by those tests could be due to non-Bayesian updating, misspecification, or a combination thereof. A natural question is then what restrictions, if any, Bayesian updating imposes on the dynamics of beliefs in isolation.

This paper provides an answer. An agent is a \emph{misspecified Bayesian} if she updates her belief using Bayes' rule given an internally consistent but possibly misspecified model of the data she observes. The paper shows that a belief sequence is consistent with misspecified Bayesianism if and only if the set of posteriors admits a partition such that, within each cell of the partition, the prior contains a ``grain'' of the corresponding conditional average posterior. This condition weakens the Bayes plausibility condition \citep*{KamenicaGentzkow2011} that characterizes the behavior of a correctly specified Bayesian. While Bayes plausibility requires the prior to equal the average posterior, the grain condition requires the prior to be a mixture of the conditional average posterior and another probability distribution on each cell of a partition.

The empirical content of this characterization is highly dependent on the environment. When the state space is finite, misspecified Bayesianism imposes only a support inclusion requirement: The support of the average posterior must be contained in the support of the prior. If the prior has full support on a finite space, any distribution of posteriors can be rationalized. When the state space is compact and the prior and average posterior admit continuous densities, the same support restriction is again both necessary and sufficient under mild regularity conditions. 

The picture changes sharply if the state space is unbounded. There the grain condition rules out posterior beliefs whose tails are heavier than those of the prior. In the Gaussian case, the restriction becomes especially sharp: If the prior is normal and every posterior realization is normal with a deterministic covariance matrix, then posterior uncertainty cannot be larger than prior uncertainty. Moreover, any movement in the mean as a result of updating must be accompanied by a strict reduction in uncertainty. Thus, in the Gaussian setting, predictable movements in the \emph{belief mean} are not by themselves a violation of Bayesian updating. Rather, rejecting Bayesian updating requires a predictable increase in \emph{belief variance}.

This sharper Gaussian characterization is useful in applications. The paper applies it to two models of overreaction in the Gaussian setting. It shows that diagnostic expectations \citep*{bordalo2018diagnostic,bordalo2020overreaction} are consistent with misspecified Bayesianism. Although diagnostic agents violate Bayes' rule at face value, their belief sequences can be rationalized by Bayesian updating under a distorted subjective signal model. By contrast, smooth diagnostic expectations \citep*{bianchi2024smooth} are inconsistent with misspecified Bayesianism under parameterizations in which the posterior variance is larger than the prior variance. The comparison illustrates a broader point of the paper. While some non-Bayesian updating rules are observationally equivalent to Bayesian updating under misspecification, others are not. The paper's characterization result delineates the exact boundary between those two cases. 

The paper's findings may appear at odds with existing results in the literature. \cite*{KamenicaGentzkow2011} show that Bayesian updating requires the average posterior to equal the prior. \citet*{Shmaya2016} show that any belief sequence in which the prior is in the relative interior of the convex hull of posteriors is consistent with Bayes' rule. This paper presents two additional theorems that clarify the relationship between these conditions and the characterization developed here. The theorems adapt the existing results to general state spaces, thus making them directly comparable to this paper's results. More importantly, they clarify that the earlier results characterize Bayesian updating only under additional restrictions on agents' subjective models. \cite*{KamenicaGentzkow2011} do so by requiring agents to have correct beliefs about the distribution of signals, whereas \cite*{Shmaya2016} require the subjective signal distribution to have the same support as the true distribution.

\paragraph*{Other related work.} The paper's results bridge two strands of work on departures from rational expectations. The first preserves Bayesian updating but allows misspecification, e.g., \citet*{esponda2016berk, esponda2021equilibrium}, \citet*{bohren2016informational}, \citet*{frick2020misinterpreting}, \citet*{fudenberg2021limit}, and \citet*{REStud,bonds}. The second posits explicitly non-Bayesian heuristics, e.g., \citet*{tversky1974judgment}, \citet*{rabin1999first}, \citet*{epstein2010non}, and \citet*{cripps2018divisible}---see \cite{ortoleva2022alternatives} for a recent survey. The results here suggest a sharp tool for empirically disentangling these  two types of deviation from data on realized priors and posteriors alone. In contemporaneous work, \citet*{Hauster_Bohren_2021} study when a non-Bayesian updating rule can be represented as misspecification once agents' forecasts of their future beliefs are also observed. By contrast, this paper characterizes misspecified Bayesianism absent such forecast data. More recently, \cite{fudenberg2026conditionally} study when an agent's initial forecast and one-step-ahead forecast revisions admit a representation by Bayesian learning under a conditionally i.i.d. model, thereby characterizing the empirical content of a much more restrictive class of subjective models than those considered here.

This paper is also related to work that tests Bayesian updating using restrictions on forecast revisions and uncertainty revisions. Most closely, \citet*{augenblick2021belief} show that, in the binary case and for a correctly specified Bayesian, expected belief movement must equal expected uncertainty reduction. This paper asks instead what remains testable once agents are allowed misspecified models of their signals. Under misspecification, Bayesian updating by itself imposes essentially no restrictions in the binary case, while a related but distinct restriction survives in the Gaussian case, namely that posterior uncertainty cannot exceed prior uncertainty and that non-trivial movements in beliefs require a strict reduction in uncertainty.

The grain condition is borrowed from \citet*{kalai1993rational}, who introduce the notion of containing a ``grain of truth'' in their study of convergence of beliefs in repeated games of incomplete information. While that literature is concerned with the long-run convergence of Bayesian learners, this paper studies the finite-horizon consistency of observed belief sequences with Bayesian updating under misspecified subjective models.

\paragraph*{Outline.} The remainder of the paper is organized as follows. Section \ref{sec:setup} introduces the conceptual framework and defines misspecified Bayesianism. Section \ref{sec:result} presents the main characterization result and explains why the partition in the theorem is necessary. Section \ref{sec:implications} studies the implications of the grain condition in several common environments. Section \ref{sec:application} applies the results to two canonical models of overreaction. Section \ref{sec:extensions} discusses extensions and the relationship to existing notions of Bayesian consistency. Section \ref{sec:conclusion} concludes. The proofs are relegated to the appendix.


\section{Conceptual framework}\label{sec:setup}

This section introduces the paper's conceptual framework and defines what it means for a belief sequence to be consistent with misspecified Bayesianism.

\subsection{Setup}
There is a fixed state of the world. The state is denoted by $x$ and belongs to a complete separable metric space $X$.\footnote{I endow $X$ with the Borel $\sigma$-algebra $\mathcal{X}\equiv \mathcal{B}(X)$ and use $\Delta(X)$ to denote the set of probability distributions over $(X,\mathcal{B}(X))$. Since $X$ is a complete separable metric space, $\Delta(X)$ is a complete separable metric space under the weak topology. I let $\Delta(\Delta(X))$ denote the set of probability distributions over $(\Delta(X),\mathcal{B}(\Delta(X)))$.}

The main objects of interest are a prior and a ``posterior'' about $x$.\footnote{I use the term ``posterior'' to refer to any probability distribution obtained by updating the prior after observing new information, regardless of whether it is derived from the prior via Bayes' rule.} The prior is a probability distribution over $X$ denoted by $\mu_0\in\Delta(X)$. The posterior is a random variable $\mu_1\in\Delta(X)$ with distribution $F_1\in \Delta(\Delta(X))$. The randomness of the posterior is due to its dependence on a random signal $s$, which belongs to a complete separable metric space $S$.\footnote{I endow $S$ with the Borel $\sigma$-algebra $\mathcal{S}\equiv \mathcal{B}(S)$ and use $\Delta(S)$ to denote the set of probability distributions over $(S,\mathcal{B}(S))$.} The \emph{true} distribution of signals given the fixed state of the world is denoted by $\mathbb{P}_S\in\Delta(S)$.


\subsection{Misspecified Bayesianism}

The paper's goal is to characterize the conditions under which the pair $(\mu_0,F_1)$ can be interpreted as arising from Bayesian updating under a possibly misspecified subjective model of how signals are generated. To do so, I present a notion of Bayesianism for general state spaces: Bayesianism requires a subjective joint distribution over states and signals, applicability of Bayes' rule following every contingency that can occur under the true signal distribution, and updating by Bayes' rule after such contingencies.\footnote{Although the second requirement is not essential for the paper's main finding, having a more demanding notion of Bayesianism strengthens the sufficiency part of the result by showing that the conclusion does not rely on the inapplicability of Bayes' rule after zero-probability events. It also allows the main result to be stated as an if and only if characterization.}

Formally, let $\mathbb Q\in\Delta(X\times S)$ denote a subjective distribution over states and signals. I require the true signal distribution $\mathbb P_S$ to be absolutely continuous with respect to the $S$-marginal $\mathbb Q_S$. This ensures that Bayes' rule is applicable following every contingency that can occur under the true signal distribution. When $\mathbb P_S$ has finite support, absolute continuity reduces to the requirement that $\mathbb Q_S(\{s\})>0$ whenever $\mathbb P_S(\{s\})>0$. 

I also require the posterior belief given signal $s$ to be represented by a regular conditional probability, which is a measurable kernel $s\mapsto\mathbb Q(\cdot|s)\in\Delta(X)$ satisfying 
\begin{equation}\label{eq:Bayes_rule}
\mathbb{Q}(D\times E) = \int_E \mathbb{Q}(D|s)\mathbb{Q}_S(ds)
\end{equation}
for all $D\in\mathcal X$ and $E\in\mathcal S$.\footnote{Such kernels exist when $X$ and $S$ are complete separable metric spaces; see \cite{Faden1985}. They are unique up to $\mathbb Q_S$-null sets.} Regular conditional probabilities define a mapping $\varphi_{\mathbb{Q}}:s\mapsto \mathbb{Q}(\cdot|s)$ from signals to posteriors, which is unique up to $\mathbb Q_S$-null sets. Bayesian updating given subjective distribution $\mathbb{Q}$ requires the prior to be updated using (a version of) this mapping. When $\mathbb{Q}_S(\{s\})>0$ for some $s$, then $\mathbb{Q}(D|s)=\mathbb{Q}(D\times\{s\})/\mathbb{Q}_S(\{s\})$ for any measurable subset $D$ of $X$, and $\varphi_{\mathbb{Q}}$ reduces to the usual Bayes' rule.

The mapping $\varphi_{\mathbb Q}$ determines the posterior as a function of the realized signal. The distribution of posteriors is therefore obtained by combining $\varphi_{\mathbb Q}$ with the true signal distribution $\mathbb P_S$. Given $\mathbb P_S$ and $\mathbb Q$, the posterior is distributed according to the pushforward measure $\mathbb P_S\circ\varphi_{\mathbb Q}^{-1}\in\Delta(\Delta(X))$, defined by
\begin{equation}\label{eq:population_dist}
\mathbb{P}_S\circ\varphi_{\mathbb{Q}}^{-1}(E) \equiv \mathbb{P}_S\left(\{s\in S:\varphi_{\mathbb{Q}}(s)\in E\}\right)
\end{equation}
for any measurable subset $E$ of $\Delta(X)$.\footnote{Under the assumption that $\mathbb{P}_S$ is absolutely continuous with respect to $\mathbb{Q}_S$, any zero $\mathbb{Q}_S$ measure set also has zero $\mathbb{P}_S$ measure. Therefore, changing the value of $\varphi_{\mathbb{Q}}$ on a $\mathbb{Q}_S$-zero probability set does not change the expression in \eqref{eq:population_dist}, so the value of $\mathbb{P}_S\circ\varphi_{\mathbb{Q}}^{-1}$ is independent of the version of $\varphi_{\mathbb{Q}}$ being used.} This is the observed distribution of posteriors when agents who hold subjective distribution $\mathbb{Q}$ observe signals distributed according to $\mathbb{P}_S$ and update their priors using Bayes' rule.


I can now define what it means for a belief sequence to be consistent with misspecified Bayesianism.
\begin{definition}\label{def:Bayesianism}
Given the true distribution of signals $\mathbb{P}_S\in\Delta(S)$, a pair $(\mu_0,F_1)$, consisting of a prior and a distribution of posteriors, is \emph{consistent with misspecified Bayesianism} if there exists a subjective distribution $\mathbb{Q}\in\Delta(X\times S)$ that satisfies the following conditions:
\begin{enumerate}[(a)]
    \item $\mathbb{Q}_X=\mu_0$,
    \item $\mathbb{P}_S$ is absolutely continuous with respect to $\mathbb{Q}_S$,
    \item $\mathbb{P}_S\circ\varphi_{\mathbb{Q}}^{-1}=F_1$,    
\end{enumerate}
where $\mathbb{Q}_X$ and $\mathbb{Q}_S$ are the $X$- and $S$-marginals of the subjective distribution $\mathbb{Q}$, respectively, $\varphi_{\mathbb{Q}}$ is the Bayesian update given subjective distribution $\mathbb{Q}$, and $\mathbb{P}_S\circ\varphi_{\mathbb{Q}}^{-1}$ is defined in \eqref{eq:population_dist}.
\end{definition}

A minimal requirement for $(\mu_0,F_1)$ to be consistent with Bayesianism given $\mathbb{P}_S$ is that $F_1=\mathbb{P}_S\circ\varphi^{-1}$ for \emph{some} measurable mapping $\varphi:S\to\Delta(X)$. If no such $\varphi$ existed, no updating rule---Bayesian or non-Bayesian---could generate a distribution $F_1$ of posteriors based on a signal distributed according to $\mathbb{P}_S$. To rule out such cases, in the remainder of the paper, I assume that $F_1$ and $\mathbb{P}_S$ can be linked via $F_1=\mathbb{P}_S\circ\varphi^{-1}$ for some measurable, possibly unknown, mapping $\varphi$. I also assume without loss of generality that $S=\Delta(X)$ (by relabeling signals by the posterior that they induce). The question is then whether $F_1=\mathbb{P}_S\circ\varphi_{\mathbb{Q}}^{-1}$ for an updating rule $\varphi_{\mathbb{Q}}$ that is Bayesian given a subjective distribution $\mathbb{Q}$ whose $X$-marginal agrees with the observed prior $\mu_0$.


\section{Characterization}\label{sec:result}
This section presents the paper's main characterization result.


\subsection{Main result}

Before presenting the main result, I introduce two definitions used in its statement.
\begin{definition}\label{def:post_mean}
Given a measurable subset $E$ of the set of posteriors $\Delta(X)$ with $F_1(E)>0$, the \emph{conditional average posterior given $E$} is the probability distribution over $X$ defined as
\[
\overline{\mu}_1^E(D) \equiv \displaystyle\frac{\int_E\mu(D)F_1(d\mu)}{F_1(E)}
\]
for any measurable set $D$. I refer to $\overline{\mu}_1\equiv \overline{\mu}_1^{\Delta(X)}$ as the \emph{average posterior}.
\end{definition}

\begin{definition}[\citealp*{kalai1993rational}]\label{def:grain}
For probability distributions $P$ and $Q$ defined over the same measurable space, $P$ \emph{contains a grain of} $Q$ if $P=\epsilon Q + (1-\epsilon) Q'$ for some $\epsilon\in (0,1]$ and some probability measure $Q'$.
\end{definition}
The following proposition states two alternative definitions, which are equivalent to Definition \ref{def:grain}:
\begin{proposition}\label{prop:unif_abs_cont}
For probability distributions $P$ and $Q$ defined over the same measurable space, the following are equivalent:
\begin{enumerate}[(i)]
    \item $P$ contains a grain of $Q$.
    \item The Radon--Nikodym derivative $f\equiv \frac{dQ}{dP}$ exists and is bounded $P$-almost surely.
    \item There exists a constant $c\geq 1$ such that $Q(E)\leq cP(E)$ for any measurable set $E$.
\end{enumerate}
\end{proposition}

The proposition illustrates that the grain condition is stronger than absolute continuity. $Q$ is absolutely continuous with respect to $P$ if $Q(E)=0$ for any event $E$ for which $P(E)=0$, whereas $P$ contains a grain of $Q$ when the ratio $Q(E)/P(E)$ is bounded uniformly in $E$. The condition in Definition \ref{def:grain} can thus be seen as a form of ``uniform absolute continuity.'' 

With Definitions \ref{def:Bayesianism}--\ref{def:grain} in hand, I can state the paper's main result.
\begin{theorem}\label{thm:main}
The pair $(\mu_0,F_1)$ is consistent with misspecified Bayesianism \emph{if and only if} there exists a countable measurable partition of the set of posteriors $\Delta(X)$ into sets $\{E_k\}_{k\in\mathbb{N}}$ such that, for every $E_k$ with $F_1(E_k)>0$,  the prior $\mu_0$ contains a grain of the conditional average posterior $\overline{\mu}_1^{E_k}$ given $E_k$.
\end{theorem}

The theorem characterizes the empirical content of Bayesian updating once the subjective model is allowed to be misspecified. In that case, Bayesian updating is no longer characterized by the requirement that the prior equals the average posterior. Instead, the prior must contain a grain of the conditional average posterior over cells of a partition of the distribution of posteriors. Furthermore, this condition is both necessary and sufficient for consistency with misspecified Bayesianism.

The need for the partition is easiest to see in the ``only if'' direction of the theorem. The belief sequence of a Bayesian agent is a martingale under the subjective distribution. Expressing that martingale restriction under the true signal distribution requires a change of measure, which introduces the likelihood ratio $d\mathbb{P}_S/d\mathbb{Q}_S.$\footnote{Condition (b) of Definition \ref{def:Bayesianism} ensures that this Radon--Nikodym derivative exists. If Condition (b) were to be dropped, the grain condition of Theorem \ref{thm:main} would still be sufficient for consistency with misspecified Bayesianism, but no longer necessary.} If this likelihood ratio is bounded, the martingale condition under the subjective distribution implies the grain condition under the true distribution. However, the likelihood ratio need not be bounded globally. The countable partition in Theorem \ref{thm:main} groups posteriors into cells on which the likelihood ratio is bounded. When the distribution of posteriors is supported on a finite set, the likelihood ratio is globally bounded, and misspecified Bayesianism is characterized by the simpler conditions stated in the next result.

\begin{proposition}\label{prop:main_finite}
Consider a pair $(\mu_0,F_1)$, with $F_1$ supported on a finite set. The following are equivalent:
\begin{enumerate}[(i)]
    \item The pair $(\mu_0,F_1)$ is consistent with misspecified Bayesianism.
    \item The prior $\mu_0$ contains a grain of any posterior $\mu_1$ in the support of $F_1$.
    \item The prior $\mu_0$ contains a grain of the average posterior $\overline{\mu}_1\equiv \int \mu F_1(d\mu)$.
\end{enumerate}
\end{proposition}
Proposition \ref{prop:main_finite} makes the link between Bayes' rule and the grain condition especially clear. If a posterior $\mu_1$ is realized with positive probability, then it must arise from conditioning on an event to which the subjective model assigns positive probability. It follows that, for any such posterior $\mu_1$, there exists a constant $c$ such that $\mu_1(D)\leq c\mu_0(D)$ for all measurable $D$, so the prior $\mu_0$ contains a grain of $\mu_1$. The more surprising part of the result is that these grain conditions are also sufficient for consistency with misspecified Bayesianism.

When the support of $F_1$ is finite, the partition in Theorem \ref{thm:main} can be taken to be the trivial partition (with one cell containing the entire space) or the singleton partition (with each cell a singleton). That makes it easier to check whether a pair $(\mu_0,F_1)$ is consistent with misspecified Bayesianism. Condition (iii) is particularly useful since it only requires the knowledge of the average posterior.


\subsection{Why the partition is necessary}\label{sec:eg:delta}
I next use an example to illustrate why the simplified characterization in Proposition \ref{prop:main_finite} is invalid when the support of $F_1$ is infinite. I start with a subjective distribution $\mathbb{Q}$ and the assumption that the posterior is generated from the prior using Bayes' rule---hence the induced prior-posterior pair will be consistent with misspecified Bayesianism by construction. I then argue that the prior does not contain a grain of any realization of the posterior or the (unconditional) average posterior. However, there exists a countable measurable partition of the set of posteriors such that the prior contains a grain of the conditional average posterior given every cell of the partition.

The state belongs to the real line: $X=\mathbb{R}$. According to the subjective distribution $\mathbb Q$, the prior distribution of the state is standard normal and the signal is identified with the state, so observing the signal fully reveals the state and the posterior is a point mass at the realized state. The true signal distribution $\mathbb{P}_S$ is as follows: The support of $\mathbb{P}_S$ coincides with the support of the $S$-marginal $\mathbb{Q}_S$ of the subjective distribution. However, the true signal distribution is $\text{Laplace}(0,1/\lambda)$, i.e., the Laplace distribution with location parameter $\mu=0$ and scale parameter $1/\lambda$. Therefore, the posterior is always a point mass at some $Z\in \mathbb{R}$, with $Z$ distributed according to the $\text{Laplace}(0,1/\lambda)$ distribution.\footnote{The example can be formalized as follows: $X=\mathbb{R}$, and $\supp \mathbb{P}_S=\supp\mathbb{Q}_S=\{\delta_x:x\in \mathbb{R}\}$, where $\delta_x$ denotes the point mass at $x$. For any measurable set $D\subset \mathbb{R}$, $\mathbb{P}_S(\{\delta_x:x\in D\})=F_{\text{Laplace}(0,1/\lambda)}(D)$ and $\mathbb{Q}_S(\{\delta_x:x\in D\})=F_{\mathcal{N}(0,1)}(D)$. Finally, $\mathbb{Q}$ is defined via equation \eqref{eq:Bayes_rule}, where $\mathbb{Q}(D|\delta_x)=\mathbb{1}\{x\in D\}$ and $\mathbb{Q}(D|s)$ is arbitrary when $s$ is not a point mass at some $x\in\mathbb{R}$. I am grateful to Eran Shmaya for suggesting this example.}

\begin{figure}[htbp]
  \centering
  \begin{minipage}[t]{0.50\textwidth}
    \centering
    $\mathcal{N}(0,1)$
    \vspace{0.3em}

    \begin{tikzpicture}[scale=1.3]
      \begin{axis}[
        width=0.9\textwidth,
        height=0.45\textwidth,
        domain=-4:4,
        samples=200,
        xtick=\empty,
        ytick=\empty,
        xlabel=\empty,
        axis lines=left,
        axis y line=none,
        xtick={0},
        xticklabels={\tiny $0$},
        axis line style = {->, >={stealth[scale=20.0]}, line width=0.7pt},
        clip=false
      ]
        \addplot[very thick, color=BrightEmerald]
          {1/sqrt(2*pi)*exp(-x^2/2)};
      \end{axis}
    \end{tikzpicture}
  \end{minipage}\hfill
  \begin{minipage}[t]{0.50\textwidth}
    \centering
    $\delta_Z, \quad Z\sim\text{Laplace}(0,1/\lambda)$
    \vspace{0.3em}

    \begin{tikzpicture}[scale=1.3]
      \begin{axis}[
        width=0.9\textwidth,
        height=0.45\textwidth,
        domain=-4:4,
        samples=200,
        xtick=\empty,
        ytick=\empty,
        xlabel=\empty,
        axis lines=left,
        axis y line=none,
        xtick={0},
        xticklabels={\tiny $0$},
        axis line style = {->, >={stealth[scale=20.0]}, line width=0.7pt},
        clip=false
      ]
        \addplot[very thick, color=Raspberry, dashed]
          {0.5*exp(-abs(x))};
        \end{axis}

        \draw[very thick, color=CobaltBlue] (0.8,0) -- (0.8,1);
        \fill[color=CobaltBlue] (0.8,1) circle[radius=2pt];
        \draw[very thick, color=CobaltBlue] (2,0) -- (2,1);
        \fill[color=CobaltBlue] (2,1) circle[radius=2pt];
        \draw[very thick, color=CobaltBlue] (4,0) -- (4,1);
        \fill[color=CobaltBlue] (4,1) circle[radius=2pt];
    \end{tikzpicture}
  \end{minipage}
  \caption{The prior (left) and posterior (right). Each point mass in the right panel is a realization of the posterior; the dashed line shows the density of the location of those point masses.}
\end{figure}

The Radon--Nikodym derivative of the $\text{Laplace}(0,1/\lambda)$ distribution with respect to the normal distribution is unbounded. Therefore, the prior does \emph{not} contain a grain of the average posterior, and statement (iii) in Proposition \ref{prop:main_finite} does not hold. Similarly, the prior does not contain a grain of any of the posteriors in the support of the distribution of posteriors, so statement (ii) in Proposition \ref{prop:main_finite} does not hold either. Yet, by construction, the induced $(\mu_0,F_1)$ pair is consistent with Bayesianism and must therefore satisfy the grain condition in Theorem \ref{thm:main}. 

The distribution of posteriors indeed satisfies the partition version of the grain condition. To see this, consider a partition of the reals into a countable union of non-empty intervals $D_k$ of finite length, and consider the measurable partition of the support of the distribution of posteriors into sets $E_k\equiv\{\delta_x:x\in D_k\}$. The average posterior over set $E_k$ is equal to the truncated distribution obtained from restricting the $\text{Laplace}(0,1/\lambda)$ distribution to $D_k$. Since the Radon--Nikodym derivative of the truncated $\text{Laplace}(0,1/\lambda)$ distribution with respect to the standard normal distribution is bounded, the prior contains a grain of the conditional average posterior given any $E_k$. 

\begin{figure}[htbp]
  \centering
  \begin{minipage}[t]{0.50\textwidth}
    \centering
    $\mathcal{N}(0,1)$
    \vspace{0.3em}

    \begin{tikzpicture}[scale=1.3]
      \begin{axis}[
        width=0.9\textwidth,
        height=0.45\textwidth,
        domain=-4:4,
        samples=200,
        xtick=\empty,
        ytick=\empty,
        xlabel=\empty,
        axis lines=left,
        axis y line=none,
        xtick={0},
        xticklabels={\tiny $0$},
        axis line style = {->, >={stealth[scale=20.0]}, line width=0.7pt},
        clip=false
      ]
        \addplot[very thick, color=BrightEmerald]
          {1/sqrt(2*pi)*exp(-x^2/2)};
      \end{axis}
    \end{tikzpicture}
  \end{minipage}\hfill
  \begin{minipage}[t]{0.50\textwidth}
    \centering
    $\text{Truncated Laplace}(0,1/\lambda)$
    \vspace{0.3em}

    \begin{tikzpicture}[scale=1.3]
      \begin{axis}[
        width=0.9\textwidth,
        height=0.45\textwidth,
        domain=-4:4,
        samples=200,
        xtick=\empty,
        ytick=\empty,
        xlabel=\empty,
        axis lines=left,
        axis y line=none,
        xtick={0},
        xticklabels={\tiny $0$},
        axis line style = {->, >={stealth[scale=20.0]}, line width=0.7pt},
        clip=false
      ]
        \addplot[very thick, color=Raspberry, dashed]
          {0.5*exp(-abs(x))};

        \addplot[very thick,color=CobaltBlue,domain=1:2]
            {4.303*exp(-x)};

        \draw[dashed, very thick, color=CobaltBlue] (axis cs:1,0) -- (axis cs:1,1.581);
        \draw[dashed, very thick, color=CobaltBlue] (axis cs:2,0) -- (axis cs:2,0.581);

        \draw[line width=0.8mm, color=black] (axis cs:1,0) -- (axis cs:2,0);
        \end{axis}
    \end{tikzpicture}
  \end{minipage}
  \caption{The prior (left) and the conditional average posterior given a positive-$F_1$-measure cell of the partition of $\Delta(X)$ (right).}
\end{figure}


\section{Implications of the grain condition}\label{sec:implications}
Theorem \ref{thm:main} characterizes misspecified Bayesianism for arbitrary state spaces. I now turn to several special cases and show that the grain condition manifests itself very differently across them. 


\subsection{Finite state spaces}
I start with the special case where the state space $X$ is finite, a common situation in applications. The following result establishes an easy-to-check condition that characterizes consistency with misspecified Bayesianism in that case:
\begin{proposition}\label{prop:finite_support}
Suppose the state space $X$ is finite. Then the pair $(\mu_0,F_1)$ is consistent with misspecified Bayesianism \emph{if and only if} 
\[
\supp\overline{\mu}_1\subseteq \supp\mu_0,
\]
where $\overline{\mu}_1\equiv \int \mu F_1(d\mu)$ denotes the average posterior.
\end{proposition}

This result characterizes misspecified Bayesianism in discrete settings. The average posterior cannot assign positive probability to states that have zero probability according to the prior. The necessity of this property is apparent given Bayes' rule; the proposition goes a step further by establishing its sufficiency. Appendix \ref{sec:discrete_example} uses a simple example to illustrate the construction used in the proof of the sufficiency part.

A corollary of Proposition \ref{prop:finite_support} is the following:
\begin{corollary}\label{cor:finite_full_support}
Suppose the state space $X$ is finite and $\mu_0$ has full support over $X$. Then the pair $(\mu_0,F_1)$ is consistent with misspecified Bayesianism for \emph{any} distribution $F_1$ of posteriors.
\end{corollary}
Misspecified Bayesianism imposes \emph{no} restrictions on posteriors when the state space is finite and the prior assigns positive probability to every state. The result casts doubt on the possibility of deciding whether decision makers are Bayesian in many common scenarios. 


\subsection{Compact state spaces}
Misspecified Bayesianism continues to impose only weak restrictions when the state space is infinite but compact and the prior and average posterior have well-behaved densities.

\begin{proposition}
\label{prop:compact}
Let the state space $X$ be a compact subset of $\mathbb{R}^n$. Suppose $\mu_0$ and $\overline{\mu}_1$ admit continuous densities $m_0$ and $\overline{m}_1$ with respect to the Lebesgue measure such that $m_0(x)>0$ for every $x\in\supp\mu_0$. Then the pair $(\mu_0,F_1)$ is consistent with misspecified Bayesianism \emph{if and only if} 
\[
\supp\overline{\mu}_1\subseteq \supp\mu_0,
\]
where $\overline{\mu}_1\equiv \int \mu F_1(d\mu)$ denotes the average posterior.
\end{proposition}
Intuitively, the uniformity in the grain condition has no bite once the state space is bounded. On a compact set and with a continuous Radon--Nikodym derivative, simple absolute continuity already implies uniform absolute continuity. Moreover, in this setting absolute continuity of $\overline{\mu}_1$ with respect to $\mu_0$ reduces to the support condition $\supp\overline{\mu}_1\subseteq \supp\mu_0$. Hence, under the regularity conditions of the proposition, misspecified Bayesianism imposes no restrictions beyond ruling out an expansion of the prior's support.


\subsection{Unbounded state spaces and tail restrictions}\label{sec:tail}
Propositions \ref{prop:finite_support} and \ref{prop:compact} suggest that misspecified Bayesianism does not impose any meaningful restrictions on belief sequences when the state space is bounded. The conclusion changes dramatically when the state space is unbounded. Then misspecified Bayesianism restricts the heaviness of posterior tails.

For concreteness, I focus on the case where the state space is $X=\mathbb{R}^n$ and use the following (partial) tail order and its uniform version:
\begin{definition}\label{def:tail_heaviness}
Let $P,Q$ be probability distributions on $\mathbb{R}^n$ with $P$ having unbounded support. If 
\[
\lim_{r\to\infty}\frac{Q(\Vert x\Vert >r)}{P(\Vert x\Vert >r)}=\infty,
\]
then $Q$ has \emph{heavier tails} than $P$.\footnote{The definition is independent of the choice of norm because of the equivalence of norms on $\mathbb{R}^n$.} 
\end{definition}

\begin{definition}\label{def:uniform_tail_heaviness}
Let $\mathcal{Q}$ be a set of probability distributions and $P$ be a probability distribution with unbounded support on $\mathbb{R}^n$. If 
\[
\lim_{r\to\infty}\inf_{Q\in \mathcal{Q}}\frac{Q(\Vert x\Vert >r)}{P(\Vert x\Vert >r)}=\infty,
\]
then distributions in $\mathcal{Q}$ have \emph{uniformly heavier tails} than $P$.
\end{definition}

These notions of tail heaviness are stronger than the log-tail comparisons often used in probability theory. They compare tail probabilities in levels rather than on a logarithmic scale. For instance, under Definition \ref{def:tail_heaviness}, a more dispersed normal distribution has heavier tails than a less dispersed one, even though the two distributions have the same tail exponent in the usual logarithmic sense. This is the relevant notion here because the grain condition is equivalent to a uniform bound on ratios of probabilities, not on ratios of log-probabilities.

The next result shows that Bayesian updating \emph{cannot} lead to tails that are uniformly heavier in the sense of Definition \ref{def:uniform_tail_heaviness}:
\begin{proposition}\label{prop:tail}
Suppose the state space is given by $X=\mathbb{R}^n$ and the prior $\mu_0$ has unbounded support. If there exists an $F_1$-positive measure set of posteriors whose tails are uniformly heavier than $\mu_0$,
then $(\mu_0,F_1)$ is \emph{not} consistent with misspecified Bayesianism.
\end{proposition}
If the support of the distribution of posteriors is finite, we can dispense with uniformity and get the following sharper result:
\begin{corollary}\label{cor:tail}
Suppose the state space is given by $X=\mathbb{R}^n$ and the prior has unbounded support. Consider a pair $(\mu_0,F_1)$, with $F_1$ supported on a finite set. If there exists a posterior $\mu_1\in \supp F_1$ whose tails are heavier than those of $\mu_0$, then $(\mu_0,F_1)$ is \emph{not} consistent with misspecified Bayesianism.
\end{corollary}

Bayesian updating can redistribute the prior's probability mass. However, shifting mass into the tail regions requires signals that were themselves extremely unlikely under the prior. Allowing for misspecified beliefs about signal probabilities enlarges the set of attainable posteriors, because signals that are rare in reality may be deemed likely by the subjective model. Yet even under such misspecification there is a hard limit. Posteriors that are (uniformly) heavier-tailed than the prior cannot be produced by Bayesian updating---no matter how misspecified the subjective model. Observing such posteriors is thus a telltale sign of violations of Bayes' rule.

\begin{example}\label{eg:tail}

The state space is the real line: $X=\mathbb{R}$. The prior is normal with mean zero and unit variance. With one-half probability, the posterior is the exponential distribution with mean $1/\lambda$; with the remaining one-half probability, the posterior is the (mirrored) exponential distribution supported over $(-\infty,0]$ with mean $-1/\lambda$. Since the prior has a strictly positive density on $\mathbb R$, both realized posteriors are absolutely continuous with respect to the prior. However, the prior does \emph{not} contain a grain of the average posterior (or either of the two realizations of the posterior), because the two posteriors in the support of $F_1$ both have heavier tails than the prior regardless of the value of $\lambda$. Therefore, Corollary \ref{cor:tail} implies that $(\mu_0,F_1)$ is inconsistent with misspecified Bayesianism---even though each realized posterior has smaller variance than the prior when $\lambda>1$. This example illustrates that, in general, shrinkage in variance, or in any other finite moment of beliefs, is neither required by Bayes' rule nor sufficient for consistency with it.
\end{example}


\subsection{Gaussian priors and Gaussian posteriors}\label{sec:gaussian}
The grain condition has a particularly simple representation in the Gaussian case. Suppose the state space is $X=\mathbb{R}^n$, the prior is multivariate normal with mean $m_0\in\mathbb{R}^n$ and positive definite covariance matrix $\Sigma_0$, and every realization of the posterior is also multivariate normal with a deterministic covariance matrix $\Sigma_1$ and a random mean $m_1\in\mathbb{R}^n$. Formally,
\[
\mu_0=\mathcal N(m_0,\Sigma_0),
\]
for some positive definite matrix $\Sigma_0 \in \mathbb{R}^{n\times n}$ and some vector $m_0\in \mathbb{R}^{n}$, and there exists some positive definite (and non-random) $\Sigma_1 \in \mathbb{R}^{n\times n}$ such that
\[
F_1(\{\mu_1: \mu_1=\mathcal N(m_1,\Sigma_1)\;\text{for some}\;m_1\in \mathbb{R}^n\})=1. 
\]

The following result characterizes when such a pair $(\mu_0,F_1)$ is consistent with misspecified Bayesianism:

\begin{theorem}\label{thm:normal}
Let $\Sigma_0$ and $\Sigma_1$ be positive definite matrices, and suppose the prior is $\mu_0=\mathcal N(m_0,\Sigma_0)$ while the posterior is $\mu_1=\mathcal N(m_1,\Sigma_1)$, where $m_1$ is a random vector in $\mathbb R^n$. Then $(\mu_0,F_1)$ is consistent with misspecified Bayesianism if and only if (i) $\Sigma_0-\Sigma_1$ is positive semidefinite and (ii) $m_1-m_0$ is in the image of $\Sigma_0-\Sigma_1$ for $F_1$-almost all posteriors.
\end{theorem}

Condition (i) states that posterior uncertainty cannot rise as a result of Bayesian updating. Condition (ii) says that if uncertainty is unchanged in some direction, then the posterior mean cannot move in that direction. Thus, misspecified Bayesianism allows arbitrary movements in the posterior mean only along directions in which the posterior variance is strictly smaller than the prior variance.

Two immediate corollaries are worth noting. First, if $\Sigma_0-\Sigma_1$ is positive definite, then condition (ii) is vacuous, so \emph{any} distribution of posterior means is consistent with misspecified Bayesianism. Second, if $\Sigma_0=\Sigma_1$, then consistency requires $m_1=m_0$ almost surely. In particular, non-trivial updating with a deterministic posterior covariance is impossible unless the posterior covariance is smaller than the prior covariance in at least one direction.

The univariate Gaussian case is the focus of much of the literature. In that case, the characterization simplifies even further.
\begin{corollary}
Suppose the prior is $\mu_0=\mathcal N(m_0,\sigma^2_0)$ while the posterior is $\mu_1=\mathcal N(m_1,\sigma^2_1)$, where $m_1$ is a random scalar and $\sigma^2_1$ is deterministic. Then $(\mu_0,F_1)$ is consistent with misspecified Bayesianism if and only if $\sigma^2_1 \leq \sigma^2_0$ with the inequality strict unless $m_1=m_0$ for $F_1$-almost all posteriors.  
\end{corollary}
Non-trivial Bayesian updating in the univariate Gaussian case requires a strict reduction in uncertainty. This restriction is related to but distinct from the restrictions derived by \citet*{augenblick2021belief} in the binary case. They show that expected belief movement must equal expected uncertainty reduction under correct specification. The restrictions here are weaker but robust to misspecification. While movements in beliefs are not by themselves informative about Bayesian updating, posterior uncertainty still cannot rise, and non-trivial updating still requires a strict reduction in uncertainty.


\section{Application: Overreaction in macroeconomic expectations}\label{sec:application}
The paper's results suggest that some non-Bayesian updating rules can be rationalized as Bayesian updating under misspecification, whereas other updating rules admit no such rationalization. Furthermore, the results delineate the boundary between those two classes of updating rules. This section illustrates these points using two leading models of overreaction in macroeconomic expectations. I show that diagnostic expectations are consistent with misspecified Bayesianism, whereas smooth diagnostic expectations, in general, are not.


\subsection{Diagnostic expectations}
I first consider the diagnostic expectations of \citet*{bordalo2020overreaction}. The authors provide evidence that macroeconomic expectations of professional forecasters overreact to news and show that this overreaction is consistent with a diagnostic updating rule in which representative states are overweighted. 

A simplified version of their environment is as follows. There is a fixed state $x\in \mathbb{R}$ and a signal $s\in\mathbb{R}$ about the state.\footnote{In \citet{bordalo2020overreaction}, the state is time-varying, and the prior itself is obtained from Kalman filtering. However, the distortion from diagnostic expectations appears in the updating step, which is the object of interest here.} The true signal distribution is given by
\begin{equation}
 s = x + \epsilon,\qquad \epsilon\sim\mathcal{N}(0,\sigma_\epsilon^2).
\label{eq:P-DGP}
\end{equation}
Agents' prior is given by
\[
\mu_0 = \mathcal{N}(m_0,\sigma_0^2).
\]
A correctly specified Bayesian then has the posterior belief 
\[
\mu^*_1=\mathcal{N}\left(m_1^*,{\sigma_1^*}^2\right),
\]
where $m^*_1=m_0 + K(s - m_0)$, ${\sigma_1^*}^2=(1-K)\sigma_0^2$, and $K\equiv \sigma_0^2/(\sigma_0^2 + \sigma_\epsilon^2)$. A diagnostic agent starts with the same prior but updates by overweighting representative states. In particular, for an agent with diagnosticity parameter $\theta$, the posterior is given by
\[
\mu^{\text{DE}}_1=\mathcal{N}\left(\tilde{m}_1,{\sigma_1^*}^2\right),
\]
where 
\[
\tilde{m}_1 = m_1^* + \theta(m^*_1 - m_0).
\]
Thus, diagnostic agents have the same posterior variance as correctly specified Bayesian agents, but their posterior mean shifts more strongly in response to new information.

The Gaussian characterization in Theorem \ref{thm:normal} can be applied directly to determine whether diagnostic expectations are misspecified Bayesian. The prior is Gaussian, every posterior realization is Gaussian, and the posterior variance is deterministic and strictly smaller than the prior variance whenever $\sigma_\epsilon^2>0$. Therefore, the conditions in Theorem \ref{thm:normal} are automatically satisfied. It follows that the belief sequence generated by diagnostic expectations is consistent with misspecified Bayesianism.

The following subjective model provides one rationalization. Let $\mathbb{Q}_X=\mu_0$, and let the conditional signal distribution under the subjective model be
\begin{equation}
 s = x + \epsilon,\qquad
 \epsilon\sim\mathcal{N}\left(\frac{-\theta}{1+\theta}(x-m_0),\frac{\sigma_\epsilon^2}{(1+\theta)^2}\right).
\label{eq:Q-conditional_DGP}
\end{equation}
This subjective model understates the noise variance and allows the observation noise to be negatively correlated with the state. Define the joint subjective distribution $\mathbb{Q}\in\Delta(X\times S)$ by
\begin{equation}
\mathbb{Q}(D\times E) \equiv \int_D \mathbb{Q}(E|x)\mu_0(dx)
\label{eq:Q-DGP}
\end{equation}
for measurable sets $D\subseteq X$ and $E\subseteq S$. Under the true signal distribution $\mathbb{P}_S$, Bayesian updating with subjective model $\mathbb{Q}$ generates the same distribution of posteriors as diagnostic expectations with parameter $\theta$.\footnote{Indeed, a stronger equivalence holds. If equation \eqref{eq:P-DGP} is interpreted as the true conditional signal distribution $\mathbb{P}(\cdot|x)$ conditional on the state $x$, then the Bayesian update $\varphi_{\mathbb{Q}}$ induced by \eqref{eq:Q-conditional_DGP} and \eqref{eq:Q-DGP} coincides with the diagnostic update $\varphi$ for every signal realization and conditional on every state. Therefore, the joint distribution of signal-posterior pairs is also matched state by state. See Subsections \ref{sec:extension-x} and \ref{sec:extension-joint}.}

This rationalization offers an alternative interpretation of overreaction in macroeconomic expectations as arising from misspecified signal models. However, the point is not that this is the correct account of overreaction, nor that the existence of a Bayesian rationalization establishes that survey respondents update as Bayesians. Rather, the point is that overreaction of the type observed in survey data and commonly explained by diagnostic expectations is not by itself inconsistent with Bayesian updating---as long as agents are allowed to entertain misspecified subjective models.


\subsection{Smooth diagnostic expectations}

I next consider smooth diagnostic expectations of \citet*{bianchi2024smooth}. Relative to diagnostic expectations, the key difference is that the distortion affects not only the posterior mean but also the posterior variance. 

The following is a simplified version of smooth diagnostic expectations in the Gaussian environment. Agents' prior is given by $\mu_0 = \mathcal{N}(m_0,\sigma_0^2)$, while the correctly specified Bayesian posterior is given by $\mu^*_1=\mathcal{N}(m_1^*,{\sigma_1^*}^2)$. There exists a reference distribution $\mathcal{N}(m_\text{ref},\sigma_\text{ref}^2)$, which need not coincide with the agent's prior and is used to measure the representativeness of any event. Smooth diagnostic expectations move beliefs away from the reference distribution and toward the Bayesian posterior, but they do so excessively. To simplify the notation, I write the model in terms of an effective diagnosticity parameter. For an agent with effective diagnosticity parameter $\theta$, the posterior is given by
\[
\mu^{\text{smooth DE}}_1=\mathcal{N}\left(\hat{m}_1,{\hat{\sigma}_1}^2\right),
\]
where 
\begin{align*}
& \hat{m}_1 = m_1^* + \theta(m^*_1 - m_\text{ref}),\\
& \hat{\sigma}_1^2 = \left(1-\frac{1-R}{R}\theta\right){\sigma_1^*}^2,
\end{align*}
and $R \equiv {\sigma_1^*}^2/\sigma_\text{ref}^2$ is the ratio of the correctly specified Bayesian variance to the variance under the reference distribution. Like diagnostic expectations, the posterior mean moves excessively in the direction of the Bayesian posterior relative to the reference distribution; unlike diagnostic expectations, the posterior variance also differs systematically from the correctly specified Bayesian posterior variance.

This difference can be critical for consistency with misspecified Bayesianism in light of Theorem \ref{thm:normal}. In the Gaussian environment, misspecified Bayesianism requires the posterior variance to be weakly smaller than the prior variance. Therefore, any parameterization of smooth diagnostic expectations satisfying
\[
\left(1-\frac{1-R}{R}\theta\right){\sigma_1^*}^2>\sigma_0^2
\]
is inconsistent with misspecified Bayesianism. Since the reference distribution need not coincide with the prior, this condition can hold for some specifications of the reference distribution and the effective diagnosticity parameter---namely, when $R>1$ and $\theta$ is sufficiently large. It follows that smooth diagnostic expectations are not, in general, observationally equivalent to misspecified Bayesianism.


\section{Extensions and alternative notions}\label{sec:extensions}
This section introduces two extensions and discusses how this paper's characterization compares with existing results in the literature.


\subsection{Rationalizing belief sequences across different states}\label{sec:extension-x} In the baseline model, the distribution of posteriors is observed given a fixed state $x$. I first consider a generalization where the distribution of posteriors is observed given every possible realization of the state. Let $\mathbb{P}\in\Delta(X\times S)$ denote the true joint distribution of the state and signal. When the realized state is $x$, signals are drawn from $\mathbb{P}(\cdot|x)$, where $\mathbb{P}(\cdot|\cdot)$ is a regular conditional probability of $\mathbb{P}$ given the Borel sigma-algebra on $X$. I let $\mathbb{P}_X$ and $\mathbb{P}_S$ denote the $X$- and $S$-marginals of $\mathbb{P}$, respectively.

The goal is to rationalize the pair $(\mu_0,\{F_{1x}\}_{x\in X})$, where $F_{1x}$ is the distribution of posteriors when the realized state of the world is $x$. The beliefs are updated using the same mapping $\varphi:S\to \Delta(X)$ regardless of the realized state of the world; that is, $F_{1x} = \mathbb{P}(\cdot|x) \circ \varphi^{-1}$ for some unknown mapping $\varphi$ and for all $x\in X$. This restriction captures the idea that the signal summarizes all the information that can be used to update beliefs. 

The following definition extends the notion of consistency with Bayesianism introduced in Definition \ref{def:Bayesianism} to this setting:
\begin{definition}\label{def:Bayesianism-x}
Given the true joint distribution of the state and signal $\mathbb{P}\in\Delta(X\times S)$, the pair $(\mu_0,\{F_{1x}\}_{x\in X})$, consisting of a prior and a distribution of posteriors in each state of the world, is \emph{consistent with misspecified Bayesianism} if there exists a subjective distribution $\mathbb{Q}\in\Delta(X\times S)$ that satisfies the following conditions:
\begin{enumerate}[(a)]
    \item $\mathbb{Q}_X=\mu_0$,
    \item $\mathbb{P}_S$ is absolutely continuous with respect to $\mathbb{Q}_S$,
    \item $\mathbb{P}(\cdot|x)\circ\varphi_{\mathbb{Q}}^{-1}=F_{1x}$ for $\mathbb{P}_X$-almost all $x$,  
\end{enumerate}
where $\varphi_{\mathbb{Q}}$ is the Bayesian update given subjective distribution $\mathbb{Q}$.
\end{definition}

Misspecified Bayesianism in this case turns out to be characterized by a condition similar to the condition that characterizes it in the fixed-state setting. The only difference is that $F_1$ in Theorem \ref{thm:main} needs to be replaced with the average of $\{F_{1x}\}_{x\in X}$ across different states. Define
\[
\overline{F}_1 \equiv \int_{X} F_{1x}\mathbb{P}_X(dx),
\]
and let
\[
\overline{\overline{\mu}}_1^E \equiv \displaystyle\frac{\int_E\mu\overline{F}_1(d\mu)}{\overline{F}_1(E)}
\]
for any measurable subset $E$ of the set of posteriors such that $\overline{F}_1(E)>0$. The following result generalizes Theorem \ref{thm:main}:

\begin{theorem}\label{thm:main-x}
The pair $(\mu_0,\{F_{1x}\}_{x\in X})$ is consistent with misspecified Bayesianism \emph{if and only if} there exists a countable measurable partition of the set of posteriors $\Delta(X)$ into sets $\{E_k\}_{k}$ such that, for every $E_k$ with $\overline{F}_1(E_k)>0$, the prior $\mu_0$ contains a grain of $\overline{\overline{\mu}}_1^{E_k}$.
\end{theorem}

Requiring state-by-state rationalization does not significantly alter the set of belief sequences that are consistent with Bayesian updating. A version of the grain condition continues to fully characterize misspecified Bayesianism. The only difference is that the grain condition is now imposed on the \emph{average} distribution of posteriors $\overline{F}_1$. When the true distribution of the state is degenerate, i.e., $\mathbb{P}_{X}(\{x^*\})=1$ for some $x^*\in X$, then Theorem \ref{thm:main-x} reduces to Theorem \ref{thm:main}. 


\subsection{Rationalizing the joint distribution of the posterior and signal}\label{sec:extension-joint}
Suppose signal realizations are observed (in addition to posteriors), and the goal is to rationalize the pair $(\mu_0,G_1)$, where $G_1\in\Delta(S\times \Delta(X))$ is the joint distribution of signal-posterior pairs. Under the assumption that the posterior is a function $\varphi$ of the realized signal, we have $G_1 = \mathbb{P}_S\circ \Phi^{-1}$, where $\mathbb{P}_S$ is the true signal distribution and the $\Phi$ mapping is defined as $\Phi:s\mapsto (s,\varphi(s))$. 

If the $\Delta(X)$-marginal of $G_1$, i.e., $F_1\equiv \mathbb{P}_S\circ \varphi^{-1}$, satisfies the grain condition of Theorem~\ref{thm:main}, then $(\mu_0,G_1)$ is consistent with misspecified Bayesianism given a subjective distribution $\mathbb{Q}$ that is identical to the one used to prove the ``if'' part of Theorem \ref{thm:main}. The reason is that the construction in Theorem \ref{thm:main} defines a mapping $\varphi_{\mathbb{Q}}$ that coincides, $\mathbb{P}_S$-almost everywhere, with the true mapping $\varphi$ used to update beliefs. Therefore, $\mathbb{P}_S\circ \Phi^{-1}=\mathbb{P}_S\circ \Phi_{\mathbb{Q}}^{-1}$, where $\Phi_{\mathbb{Q}}:s\mapsto (s,\varphi_{\mathbb{Q}}(s))$.


\subsection{Bayes plausibility and related notions}\label{sec:KG_SY}
Definition \ref{def:Bayesianism} puts no restrictions on what constitutes a reasonable subjective distribution $\mathbb{Q}$. The assumption that any well-defined subjective distribution is permissible is made in keeping with \citet*{savage1972foundations}'s idea of purely subjective probability. I assume that any subjective distribution $\mathbb{Q}$ that can rationalize $(\mu_0, F_1)$ is a valid subjective distribution. In other words, rationality of beliefs is not judged by what those beliefs are but by how they are updated. I next discuss two alternatives to this assumption proposed in the literature and how they change the conclusion of Theorem \ref{thm:main}.

The first alternative I consider imposes a correctly specified belief about the distribution of signals. This assumption leads to Bayes plausibility (or the martingale property of Bayesian beliefs): The average posterior must equal the prior. \citet*{aumann1995repeated} and \citet*{KamenicaGentzkow2011} show that this is indeed the only restriction Bayesian updating imposes on beliefs. The following theorem adapts this result to general state spaces. More importantly, however, it highlights the fact that Bayes plausibility characterizes Bayesianism \emph{only} under the assumption of correct beliefs about the distribution of signals.
\begin{theorem}[\citealp*{KamenicaGentzkow2011}]\label{thm:Kamenica_Gentzkow}
The pair $(\mu_0,F_1)$ is consistent with Bayesianism given a subjective distribution $\mathbb{Q}$ with the $S$-marginal satisfying $\mathbb{Q}_S=\mathbb{P}_S$ \emph{if and only if} $\mu_0=\overline{\mu}_1\equiv \int \mu F_1(d\mu)$.
\end{theorem}

A more permissive notion of Bayesianism is proposed by \citet*{Shmaya2016}. They allow for incorrect beliefs about the distribution of signals---as long as the supports of those beliefs coincide with the support of the true distribution. The following theorem generalizes \citet*{Shmaya2016}'s Lemma 1 to general metric state spaces and arbitrary true signal distributions. It reduces to their result when both $X$ and $\supp\mathbb{P}_S$ are finite sets. However, its main significance is to clarify that \citet*{Shmaya2016}'s conclusion relies on an a priori restriction on what constitutes a reasonable subjective distribution. 

\begin{theorem}[\citealp*{Shmaya2016}]\label{thm:Shmaya_Yariv}
The following statements are equivalent:
\begin{enumerate}[(i)]
    \item The pair $(\mu_0,F_1)$ is consistent with Bayesianism given a subjective distribution $\mathbb{Q}$ whose $S$-marginal $\mathbb{Q}_S$ is equivalent to $\mathbb{P}_S$.\footnote{Probability distributions $P$ and $Q$ are \emph{equivalent} if $Q$ is absolutely continuous with respect to $P$ and $P$ is absolutely continuous with respect to $Q$.}
    \item There exists a probability measure $\lambda \in \Delta(\Delta(X))$ such that $\lambda$ and $F_1$ are equivalent and $\mu_0=\int \mu\lambda(d\mu)$.
\end{enumerate}
\end{theorem}

Theorems \ref{thm:Kamenica_Gentzkow} and \ref{thm:Shmaya_Yariv} show that misspecified Bayesianism is a less restrictive notion than either of the notions considered by \cite{KamenicaGentzkow2011} and \cite{Shmaya2016}. The following table summarizes the relationship between different notions of Bayesianism and the conditions that characterize them:

\begin{center}
{\footnotesize
\begin{framed}
\begin{align*}
  & \text{Bayes plausibility (KG, 2011)}
    & \multirow{2}{*}{\(\implies\)} 
      \quad 
    & \text{Shmaya and Yariv (2016)}
    & \multirow{2}{*}{\(\implies\)}
      \quad 
    & \text{misspecified Bayesianism}
  \\[-0.5ex]
  & \quad\quad \mu_0=\int \mu F_1(d\mu)
    & 
    & \mu_0=\int \mu \lambda(d\mu), \quad \lambda \sim F_1
    & 
    & \mu_0 = \epsilon \int \mu F_1(d\mu) + (1-\epsilon)\mu_1'
\end{align*}
\end{framed}}
\end{center}


\section{Conclusion}\label{sec:conclusion}
The paper's theoretical results have two implications for empirical work. First, they show that many existing tests of Bayesian updating are in fact joint tests of Bayesian updating and correct specification of agents' subjective models. A rejection by such tests may therefore reflect non-Bayesian updating, misspecification, or both. More generally, the paper shows that once arbitrary misspecification is allowed, Bayesian updating, in isolation, has very little empirical content in finite or bounded state spaces.

Second, the paper shows that Bayesian updating remains falsifiable even under arbitrary misspecification and proposes ways of testing it. Doing so requires moving beyond bounded state spaces and focusing on those features of beliefs that remain informative about Bayesian updating under misspecification. In Gaussian environments, the empirical content of Bayesian updating is especially sharp: Posterior uncertainty cannot exceed prior uncertainty regardless of the extent of misspecification. An increase in the dispersion of Gaussian beliefs is therefore a signature of non-Bayesian updating that is robust to misspecification.

More broadly, the analysis shows that allowing for misspecified subjective models greatly expands the range of belief sequences that are consistent with Bayes' rule---but not without limit. The paper characterizes exactly where those limits lie.


\appendix
\section{Proofs}


\subsection*{Proof of Proposition \ref{prop:unif_abs_cont}}
The proof involves showing (i) $\implies$ (ii) $\implies$ (iii) $\implies$ (i).

\vspace{1em}
\noindent\emph{Proof of (i) $\implies$ (ii).} If $P=\epsilon Q + (1-\epsilon) Q'$, then $Q(E)\leq \frac{1}{\epsilon}P(E)$ for any measurable set $E$. Therefore, $Q$ is absolutely continuous with respect to $P$, so by the Radon--Nikodym theorem, there exists a derivative $f\equiv \frac{dQ}{dP}$. I finish the proof by showing that $f$ is bounded $P$-almost surely. Toward a contradiction, suppose that for any positive constant $C$ there exists a measurable set $E$ with $P(E)>0$ such that $f>C$ on $E$. Then,
\[
Q(E)=\int_E dQ = \int_E f dP>C\int_E dP=CP(E).
\]
Since $C$ is arbitrary, there exists no constant $\epsilon>0$ such that $Q(E)\leq \frac{1}{\epsilon}P(E)$ for all $E$, a contradiction.

\vspace{1em}
\noindent\emph{Proof of (ii) $\implies$ (iii).} 
Suppose the Radon--Nikodym derivative $f\equiv \frac{dQ}{dP}$ satisfies $f\leq c$ for some $c\geq 1$ and up to sets of zero $P$ measure. For any measurable set $E$,
\[
Q(E) = \int_E dQ = \int_E f dP \leq c \int_E dP = cP(E).
\]

\vspace{1em}
\noindent\emph{Proof of (iii) $\implies$ (i).} Let $\epsilon\equiv 1/c\leq 1$. When $\epsilon=1$, then $Q'$ can be chosen arbitrarily. This is because $Q(E)\leq P(E)$ implies $Q(E^c)\geq P(E^c)$, where $E^c$ denotes the complement of $E$. But  $Q(E^c)\leq P(E^c)$ by assumption. Therefore, $Q(E)=P(E)$. Since $E$ is an arbitrary measurable set, $Q=P$. When $\epsilon<1$, set $Q'=\frac{1}{1-\epsilon}P - \frac{\epsilon}{1-\epsilon}Q$. To finish the proof, I need to argue that such a $Q'$ is a probability measure. Note that $Q'(E)=\frac{1}{1-\epsilon}P(E) - \frac{\epsilon}{1-\epsilon}Q(E)\geq \frac{1}{1-\epsilon}P(E) - \frac{1}{1-\epsilon}P(E)=0$. Moreover, since $P$ and $Q$ are probability measures, $Q'$ returns zero for the empty set, returns one
for the entire space, and is countably additive. Therefore, $Q'$ is a probability measure.\hfill\qed


\subsection*{Proof of Theorem \ref{thm:main}}
\noindent\emph{Proof of the ``if'' direction.} The proof of this direction is constructive. Given the measurable space $(X,\mathcal{X})$ and the true signal distribution $\mathbb{P}_S$, I construct the subjective distribution $\mathbb{Q}$ that rationalizes an observed pair $\big(\mu_0,F_1\big)$ satisfying the assumption of the theorem. By assumption, $F_1=\mathbb{P}_S\circ\varphi^{-1}$ for some $\varphi$, and there exists a countable measurable partition $\{E_k\}_{k\in\mathbb{N}}$ of the set of posteriors $\Delta(X)$ into sets such that, for every $E_k$ with $F_1(E_k)>0$, $\mu_0$ contains a grain of the conditional average posterior $\overline{\mu}_1^{E_k}$ given $E_k$. Let $K$ denote the indices of the cells $E_k$ for which $F_1(E_k)>0$. For any $k\in K$, since $\mu_0$ contains a grain of $\overline{\mu}_1^{E_k}$, there exist some $\epsilon_k\in(0,1]$ and some probability measure $\mu'_k\in \Delta(X)$ such that $\mu_0=\epsilon_k\overline{\mu}_1^{E_k}+(1-\epsilon_k)\mu'_k$. Define $\hat{S}\equiv \bigcup_{k\in K} E_k$, and note that $F_1(\hat{S})=1$.

I start by constructing the regular conditional probability $\mathbb{Q}(\cdot|\cdot):\mathcal{X}\times S\to [0,1]$ that represents the posterior about state $x\in X$ conditional on signal $s$. The case in which $X$ is a singleton is trivial, so assume $X$ contains at least two points. Then $S=\Delta(X)$ is uncountable, and since there are at most countably many signals $s\in S$ such that $\mathbb{P}_S(\{s\})>0$, there exists a signal $\ominus\in S$ such that $\mathbb P_S(\{\ominus\})=0$. For any $s\in S$ such that $\varphi(s)\in\hat{S}$ and $s\neq \ominus$, set $\mathbb{Q}(D|s)=\varphi(s)(D)$ for all $D\in\mathcal{X}$. Set $\mathbb{Q}(\cdot|\ominus)=\mu'$, where $\mu'$ is an arbitrary probability measure over $X$ if $1-\sum_{k\in K}\epsilon_kF_1(E_k)=0$ and is given by 
\[
\mu'(D)\equiv \sum_{k\in K}\frac{(1-\epsilon_k)F_1(E_k)}{1-\sum_{k\in K}\epsilon_kF_1(E_k)}\mu'_k(D)
\]
for all $D\in\mathcal{X}$ if $1-\sum_{k\in K}\epsilon_kF_1(E_k)>0$. Note that $\mu'$ is always a probability measure over $X$ since $\epsilon_k\leq 1$ for all $k\in K$ and $\sum_{k\in K}\frac{(1-\epsilon_k)F_1(E_k)}{1-\sum_{k\in K}\epsilon_kF_1(E_k)}=1$. Finally, set $\mathbb{Q}(D|s)=\mu_0(D)$ for any $s\in S$ such that $s\neq \ominus$ and $\varphi(s)\notin \hat{S}$ and all $D\in\mathcal{X}$, indicating that the posterior equals the prior conditional on any signal realized with zero probability. Note that, by construction, the mapping $s\mapsto\mathbb{Q}(D|s)$ is measurable for any $D\in\mathcal{X}$, and $\mathbb{Q}(\cdot|s)$ is a probability distribution on $(X,\mathcal{X})$ for all $s\in S$. 

I can now define the subjective distribution $\mathbb{Q}$, starting with its $S$-marginal distribution $\mathbb{Q}_S$. Let 
\begin{equation}
\mathbb{Q}_S(E)\equiv\sum_{k\in K}\epsilon_k\mathbb{P}_S\left(E\cap \varphi^{-1}(E_k)\right)+\left(1-\sum_{k\in K}\epsilon_kF_1(E_k)\right)\mathbb{1}\{\ominus\in E\},\label{eq:Q_S_definition}
\end{equation}
for all $E\in\mathcal{S}$. The fact that $\epsilon_k\leq 1$ for all $k\in K$ implies that $\sum_{k\in K}\epsilon_kF_1(E_k)\leq 1$. Therefore, $\mathbb{Q}_S(E)$ is a probability distribution over $S$. Next, let
\begin{equation}\label{eq:Bayes_rule_proof}
\mathbb{Q}(D\times E) \equiv \int_E \mathbb{Q}(D|s)\mathbb{Q}_S(ds)
\end{equation}
for all $D\in\mathcal{X}$ and $E\in \mathcal{S}$. Since the sigma-algebra $(\mathcal{X}\times\mathcal{S})$ over $(X\times S)$ is generated by sets of the form $D\times E$ with $D\in\mathcal{X}$ and $E\in \mathcal{S}$, the above expression fully specifies the probability distribution $\mathbb{Q}$. Furthermore, comparing equations \eqref{eq:Bayes_rule} and \eqref{eq:Bayes_rule_proof} shows that $\mathbb{Q}(\cdot|\cdot)$ is indeed a regular conditional probability of $\mathbb{Q}$ given $\mathcal{S}$.

I next show that the true distribution $\mathbb{P}_S$ is absolutely continuous with respect to $\mathbb{Q}_S$. It suffices to show that if $\mathbb{Q}_S(E)=0$ for some $E\in\mathcal{S}$, then also $\mathbb{P}_S(E)=0$. Fix such a set $E$. Since $\mathbb{Q}_S(E)\geq \sum_{k\in K}\epsilon_k\mathbb{P}_S(E\cap \varphi^{-1}(E_k))$ and $E$ has zero $\mathbb{Q}_S$ measure, $\mathbb{P}_S(E\cap \varphi^{-1}(E_k))=0$ for all $k\in K$. Therefore, $\mathbb{P}_S(E\cap \varphi^{-1}(\hat{S}))=0$. On the other hand, $F_1(\hat{S})=1$ implies that $\mathbb{P}_S(\varphi^{-1}(\hat{S}))=1$, and so $\mathbb{P}_S(S\setminus\varphi^{-1}(\hat{S}))=0$. Therefore,
\[
\mathbb{P}_S(E) = \mathbb{P}_S(E\cap \varphi^{-1}(\hat{S}))+ \mathbb{P}_S(E\cap(S \setminus \varphi^{-1}(\hat{S})))\leq \mathbb{P}_S(E\cap \varphi^{-1}(\hat{S}))+ \mathbb{P}_S(S\setminus \varphi^{-1}(\hat{S}))=0.
\]

It remains to show that $\mathbb{Q}_X=\mu_0$ and that the distribution of posteriors $\mathbb{P}_S\circ \varphi_{\mathbb{Q}}^{-1}$, defined in equation \eqref{eq:population_dist}, coincides with the observed posterior distribution $F_1$. Equation \eqref{eq:Q_S_definition} implies that $\mathbb{Q}_S(\varphi^{-1}(\hat{S})\cup \{\ominus\})=1$. Therefore, for any $D\in\mathcal{X}$,
\begin{align*}
   \mathbb{Q}_X(D)& =\int_S \mathbb{Q}(D|s)\mathbb{Q}_S(ds)\\ & = \int_{\varphi^{-1}(\hat{S})\setminus\{\ominus\}}\varphi(s)(D)\mathbb{Q}_S(ds)+\mathbb{Q}_S(\{\ominus\})\mathbb{Q}(D|\ominus)\\ & = \sum_{k\in K}\epsilon_k\int_{\varphi^{-1}(E_k)}\varphi(s)(D)\mathbb{P}_S(ds)+\left(1-\sum_{k\in K}\epsilon_kF_1(E_k)\right)\mu'(D)\\ & = \sum_{k\in K}\epsilon_k\int_{E_k}\mu(D)\mathbb{P}_S\circ\varphi^{-1}(d\mu)+\left(1-\sum_{k\in K}\epsilon_kF_1(E_k)\right)\mu'(D)\\ & = \sum_{k\in K}\epsilon_k\int_{E_k}\mu(D) F_1(d\mu)+\left(1-\sum_{k\in K}\epsilon_kF_1(E_k)\right)\mu'(D)\\ & = \sum_{k\in K}F_1(E_k)\epsilon_k\overline{\mu}_1^{E_k}(D)+\left(1-\sum_{k\in K}\epsilon_kF_1(E_k)\right)\mu'(D)\\ & = \sum_{k\in K}F_1(E_k)\left(\mu_0(D)-(1-\epsilon_k)\mu'_k(D)\right)+\left(1-\sum_{k\in K}\epsilon_kF_1(E_k)\right)\mu'(D)\\ & = \mu_0(D)-\sum_{k\in K}(1-\epsilon_k)F_1(E_k)\mu'_k(D)+\left(1-\sum_{k\in K}\epsilon_kF_1(E_k)\right)\mu'(D).
\end{align*}
If $\epsilon_k=1$ for all $k\in K$, then the last two terms in the above display are both zero, and so $\mathbb{Q}_X(D)=\mu_0(D)$. If, on the other hand, $\epsilon_k<1$ for some $k\in K$, then $1-\sum_{k\in K}\epsilon_kF_1(E_k)>0$, and the last two terms cancel out given the definition of $\mu'$, again resulting in $\mathbb{Q}_X(D)=\mu_0(D)$.

Lastly, I show that $\mathbb{P}_S\circ\varphi_{\mathbb{Q}}^{-1}=F_1$. Since $F_1=\mathbb{P}_S\circ\varphi^{-1}$, $F_1(\hat{S})=1$, and $\mathbb{P}_S(\{\ominus\})=0$,
\[
\mathbb{P}_S(\{s\in S: \varphi(s) \in \hat{S}, s\neq \ominus\})=1.
\]
Therefore, for any $E\in\mathcal{S}$,
\begin{align*}
\mathbb{P}_S\circ \varphi_{\mathbb{Q}}^{-1}(E) & =\mathbb{P}_S\left(\{s\in S:\mathbb{Q}(\cdot|s)\in E\}\right)\\ & =\mathbb{P}_S\left(\{s\in S:\mathbb{Q}(\cdot|s)\in E,\varphi(s)\in\hat{S},s\neq \ominus\}\right)\\ & =\mathbb{P}_S\left(\{s\in S:\varphi(s)\in E\}\right)=F_1(E).
\end{align*}
This completes the proof of the first direction.

\vspace{1em}
\noindent\emph{Proof of the ``only if'' direction.} Let $\mathbb{Q}$ denote the subjective distribution on $X\times S$, and let $\mathbb{Q}(\cdot|\cdot)$ denote a regular conditional probability of $\mathbb{Q}$ given $\mathcal{S}$. Since signal labels have no inherent meaning, I can relabel each signal by the posterior belief it induces. More specifically, by the argument in the footnote below, I may assume without loss of generality that $\mathbb{Q}(D|\mu)=\mu(D)$ for any $\mu\in S=\Delta(X)$. Under this normalization, $\varphi_{\mathbb{Q}}$ is the identity mapping and $F_1=\mathbb{P}_S$.\footnote{Given a subjective distribution  $\mathbb{Q}$ on $X\times S$ with regular conditional probability $\mathbb{Q}(\cdot|\cdot)$ and a true distribution for signals $\mathbb{P}_S\in\Delta(S)$, define $\tilde{\mathbb{Q}}(D|\mu)\equiv\mu(D)$, $\tilde{\mathbb{Q}}_S(E)\equiv\mathbb{Q}_S(\{s\in S:\mathbb{Q}(\cdot|s)\in E\})$, $\tilde{\mathbb{Q}}(D\times E)=\int_E\tilde{\mathbb{Q}}(D|s)\tilde{\mathbb{Q}}_S(ds)$, and $\tilde{\mathbb{P}}_S(E)\equiv  \mathbb{P}_S(\{s\in S:\mathbb{Q}(\cdot|s)\in E\})$ for any $\mu\in S =\Delta(X)$ and any measurable sets $D\subseteq X$ and $E\subseteq S$. Then $\tilde{\mathbb{Q}}_X=\mathbb{Q}_X$ and $\tilde{\mathbb{P}}_S\circ \varphi_{\tilde{\mathbb{Q}}}^{-1}=\mathbb{P}_S\circ \varphi_{\mathbb{Q}}^{-1}$; that is, the prior and distribution of posteriors induced by $\tilde{\mathbb{Q}}$ and $\tilde{\mathbb{P}}_S$ coincide with those induced by $\mathbb{Q}$ and $\mathbb{P}_S$. Moreover, $\tilde{\mathbb{P}}_S$ is absolutely continuous with respect to $\tilde{\mathbb{Q}}_S$. The rest of the proof goes through verbatim by replacing $(\mathbb Q,\mathbb P_S)$ by $(\tilde{\mathbb Q},\tilde{\mathbb P}_S)$ if necessary.} 

Since $\mathbb{Q}$ satisfies condition (a) of Definition \ref{def:Bayesianism} and $\mathbb{Q}(\cdot|\cdot)$ is a regular conditional probability of $\mathbb{Q}$ given $\mathcal{S}$,
\begin{equation}
\mu_0(D) = \mathbb{Q}_X(D) = \int_{S}\mathbb{Q}(D|s)\mathbb{Q}_S(ds)\label{eq:onlyif1}
\end{equation}
for all $D\in\mathcal{X}$. On the other hand, by condition (b) of Definition \ref{def:Bayesianism}, $\mathbb{P}_S$ is absolutely continuous with respect to $\mathbb{Q}_S$. Hence, by the Radon--Nikodym theorem, there exists a Radon--Nikodym derivative $f\equiv \frac{d\mathbb{P}_S}{d\mathbb{Q}_S}$. For $k\in \mathbb{N}$, define
\[
E_k \equiv \big\{s\in S:f(s)\in[k-1,k)\big\}.
\]
Since $f$ is a measurable function, $E_k$ is a measurable subset of $S$ for any $k\in\mathbb{N}$. Furthermore, the sets $\{E_k\}_{k\in\mathbb N}$ form a countable measurable partition of $S=\Delta(X)$. For any $k$ such that $F_1(E_k)>0$,
\begin{equation*}
\overline{\mu}_1^{E_k}=\frac{1}{F_1(E_k)}\int_{E_k} \mu F_1(d\mu)=\frac{1}{F_1(E_k)}\int_{E_k} \mu \;\mathbb{P}_S\circ\varphi_{\mathbb{Q}}^{-1}(d\mu)=\frac{1}{F_1(E_k)}\int_{\varphi_{\mathbb{Q}}^{-1}(E_k)}\mathbb{Q}(\cdot|s)\mathbb{P}_S(ds),
\end{equation*}
where the last equality uses the change-of-variables formula for pushforward measures. Therefore, since $\varphi_{\mathbb{Q}}^{-1}(E_k)=E_k$,
\begin{equation}
\overline{\mu}_1^{E_k}(D) = \frac{1}{F_1(E_k)}\int_{E_k}\mathbb{Q}(D|s)\mathbb{P}_S(ds)\label{eq:onlyif2}    
\end{equation}
for any $D\in\mathcal{X}$.
Since $f$ is the Radon--Nikodym derivative of $\mathbb{P}_S$ with respect to $\mathbb{Q}_S$,
\begin{equation}
\int_{E_k}\mathbb{Q}(D|s)\mathbb{P}_S(ds) = \int_{E_k}\mathbb{Q}(D|s)f(s)\mathbb{Q}_S(ds) \leq k \int_{E_k}\mathbb{Q}(D|s)\mathbb{Q}_S(ds)\leq k \int_{S}\mathbb{Q}(D|s)\mathbb{Q}_S(ds),\label{eq:onlyif3}    
\end{equation}
where the first inequality is by the definition of set $E_k$, and the second inequality is due to the fact that $\int_{S\setminus E_k}\mathbb{Q}(D|s)\mathbb{Q}_S(ds)\geq 0$. Equations \eqref{eq:onlyif1}--\eqref{eq:onlyif3} imply
\[
\overline{\mu}_1^{E_k}(D) = \frac{1}{F_1(E_k)}\int_{E_k}\mathbb{Q}(D|s)\mathbb{P}_S(ds)\leq \frac{k}{F_1(E_k)}\int_{S}\mathbb{Q}(D|s)\mathbb{Q}_S(ds) = \frac{k}{F_1(E_k)}\mu_0(D).
\]
The $k/F_1(E_k)$ constant in the above inequality is independent of $D$. Therefore, by Proposition~\ref{prop:unif_abs_cont}, $\mu_0$ contains a grain of $\overline{\mu}_1^{E_k}$.\hfill\qed


\subsection*{Proof of Proposition \ref{prop:main_finite}}
Theorem \ref{thm:main} establishes (iii) $\implies$ (i) by choosing the trivial partition, so I only need to show that (i) $\implies$ (ii) $\implies$ (iii).

\vspace{1em}
\noindent\emph{Proof of (i) $\implies$ (ii).} Fix a posterior $\mu_1$ in the support of $F_1$. I show that $\mu_0$ contains a grain of $\mu_1$. Let $\mathbb{Q}$ denote the subjective distribution on $X\times S$, and let $\mathbb{Q}(\cdot|\cdot)$ denote the regular conditional probability of $\mathbb{Q}$ given $\mathcal{S}$. Let $S_{\mu_1}=\{s\in S: \mathbb{Q}(\cdot|s)=\mu_1\}$ denote the set of signals that engender $\mu_1$ as the posterior. The assumptions that the support of $F_1$ is finite and $\mu_1$ is in the support of $F_1$ imply that $F_1(\{\mu_1\})>0$, and so $\mathbb{P}_S(S_{\mu_1})>0$. Since $\mathbb{P}_S$ is absolutely continuous with respect to $\mathbb{Q}_S$, it must also be that $\mathbb{Q}_S(S_{\mu_1})>0$. Since $\mathbb{Q}(\cdot|\cdot)$ is the regular conditional probability of $\mathbb{Q}$ given $\mathcal{S}$, 
\[
\mathbb{Q}(D\times S_{\mu_1})=\int_{S_{\mu_1}}\mathbb{Q}(D|s)\mathbb{Q}_S(ds)=\mu_1(D)\mathbb{Q}_S(S_{\mu_1})
\]
for any $D\in\mathcal{X}$. On the other hand,
\[
\mu_0(D) = \mathbb{Q}_X(D)=\mathbb{Q}(D\times S)\geq \mathbb{Q}(D\times S_{\mu_1}).
\]
Therefore,
\[
\mu_1(D) =\frac{\mathbb{Q}(D\times S_{\mu_1})}{\mathbb{Q}_S(S_{\mu_1})}\leq \frac{1}{\mathbb{Q}_S(S_{\mu_1})}\mu_0(D),
\]
and by Proposition \ref{prop:unif_abs_cont}, $\mu_0$ contains a grain of $\mu_1$.

\vspace{1em}
\noindent\emph{Proof of (ii) $\implies$ (iii).}
By assumption, for any $\mu_1\in \supp F_1$, there exist some $\epsilon_{\mu_1}\in(0,1]$ and some probability measure $\mu'_{\mu_1}\in\Delta(X)$ such that $\mu_0=\epsilon_{\mu_1}\mu_1+(1-\epsilon_{\mu_1})\mu'_{\mu_1}$. Therefore,
\[
\overline{\mu}_1 = \sum_{\mu_1\in \supp F_1}F_1(\{\mu_1\})\mu_1=\sum_{\mu_1\in \supp F_1}F_1(\{\mu_1\})\left(\frac{\mu_0}{\epsilon_{\mu_1}}-\frac{1-\epsilon_{\mu_1}}{\epsilon_{\mu_1}}\mu'_{\mu_1}\right),
\]
and so
\begin{equation}\label{eq:proof_step_finite}
\mu_0 =\frac{1}{\sum_{\mu_1\in \supp F_1}\frac{F_1(\{\mu_1\})}{\epsilon_{\mu_1}}}\left(\overline{\mu}_1+\sum_{\mu_1\in \supp F_1}\frac{(1-\epsilon_{\mu_1})F_1(\{\mu_1\})}{\epsilon_{\mu_1}}\mu'_{\mu_1}\right).  
\end{equation}
Let $\epsilon\equiv\left(\sum_{\mu_1\in \supp F_1}\frac{F_1(\{\mu_1\})}{\epsilon_{\mu_1}}\right)^{-1}$ denote the weighted harmonic mean of $\{\epsilon_{\mu_1}\}_{\mu_1\in \supp F_1}$. Since all $\epsilon_{\mu_1}$ are in the $(0,1]$ interval, so is $\epsilon$. If $\epsilon=1$, choose $\mu'\in\Delta(X)$ arbitrarily. Otherwise, let
\[
\mu'\equiv \frac{\epsilon}{1-\epsilon}\sum_{\mu_1\in \supp F_1}\frac{(1-\epsilon_{\mu_1})F_1(\{\mu_1\})}{\epsilon_{\mu_1}}\mu'_{\mu_1}
\]
denote the weighted average of probability measures $\mu'_{\mu_1}$. Since the weights are positive and add up to one, $\mu'$ is a probability distribution over $X$. Equation \eqref{eq:proof_step_finite} can thus be written as
$\mu_0=\epsilon\overline{\mu}_1+(1-\epsilon)\mu'$, establishing that $\mu_0$ contains a grain of $\overline{\mu}_1$.\hfill\qed


\subsection*{Proof of Proposition \ref{prop:finite_support}}

\noindent\emph{Proof of the ``if'' direction.}
If $\supp\overline{\mu}_1\subseteq \supp\mu_0$, then
\begin{equation}
\mu_0(\{x\})=0 \implies \overline{\mu}_1(\{x\})=0.\label{eq:grain_support}    
\end{equation}
Define
\[
c \equiv \max_{\left\{x\in X: \mu_0(\{x\})>0\right\}}\frac{\overline{\mu}_1(\{x\})}{\mu_0(\{x\})}.
\]
Since $X$ is a finite set, $c$ is well-defined. Furthermore, since $\mu_0$ and $\overline{\mu}_1$ are probability distributions over $X$, which also satisfy \eqref{eq:grain_support}, $c\geq 1$. Therefore, for all $x\in X$,
\[
\overline{\mu}_1(\{x\}) \leq c \mu_0(\{x\}),
\]
where the inequality follows the definition of $c$ for any $x$ for which $\mu_0(\{x\})>0$ and follows equation \eqref{eq:grain_support} for other $x$. The above display establishes that $\mu_0$ contains a grain of $\overline{\mu}_1$. The result then follows Theorem \ref{thm:main} by choosing the trivial partition of $\Delta(X)$.

\vspace{1em}
\noindent\emph{Proof of the ``only if'' direction.} Let $\mathbb{Q}$ denote the subjective distribution on $X\times S$, and let $\mathbb{Q}(\cdot|\cdot)$ denote the regular conditional probability of $\mathbb{Q}$ given $\mathcal{S}$. By the argument in the proof of the ``only if'' direction of Theorem \ref{thm:main},
\begin{align}
& \mu_0(D) = \int_{S}\mathbb{Q}(D|s)\mathbb{Q}_S(ds),\label{eq:mu_0_star}\\
& \overline{\mu}_1(D)=\int_{S}\mathbb{Q}(D|s)\mathbb{P}_S(ds).\label{eq:mu_1_bar_star}
\end{align}
for any $D\in\mathcal{X}$. Fix some $x\in X$ \emph{not} in the support of $\mu_0$, and let $\hat{S}\equiv \{s\in S:\mathbb{Q}(\{x\}|s)>0\}$. Since $\mu_0(\{x\})=0$, by equation \eqref{eq:mu_0_star}, $\mathbb{Q}_S(\hat{S})=0$. Therefore, since $\mathbb{P}_S$ is absolutely continuous with respect to $\mathbb{Q}_S$, it must be that $\mathbb{P}_S(\hat{S})=0$. Equation \eqref{eq:mu_1_bar_star} then implies that $\overline{\mu}_1(\{x\})=0$; that is, $x$ is not in the support of $\overline{\mu}_1$.\hfill\qed


\subsection*{Proof of Proposition \ref{prop:compact}}
\noindent\emph{Proof of the ``if'' direction.} By assumption, $m_0$ and $\overline{m}_1$ are continuous functions and $m_0(x)>0$ for all $x\in \supp \mu_0$. Therefore, $\overline{m}_1(x)/m_0(x)$ is a continuous function over the compact support of $\mu_0$, and is therefore bounded. But since $\mu_0$ and $\overline{\mu}_1$ have densities, the Radon--Nikodym derivative $d\overline{\mu}_1/d\mu_0$ is equal to the ratio of densities $\overline{m}_1(x)/m_0(x)$, $\mu_0$-almost everywhere. (Since $\supp \overline{\mu}_1\subseteq \supp \mu_0$, the Radon--Nikodym derivative is arbitrary and irrelevant off the support of $\mu_0$.) Thus, Proposition \ref{prop:unif_abs_cont} implies that $\mu_0$ contains a grain of $\overline{\mu}_1$. The result then follows Theorem \ref{thm:main} by choosing the trivial partition.

\vspace{1em}
\noindent\emph{Proof of the ``only if'' direction.} Theorem \ref{thm:main} implies that there exists a measurable partition of the set of posteriors $\Delta(X)$ into sets $\{E_k\}_{k}$ such that, for every $E_k$ with $F_1(E_k)>0$,  the prior $\mu_0$ contains a grain of the conditional average posterior $\overline{\mu}_1^{E_k}$ given $E_k$. Therefore, for every $E_k$ with $F_1(E_k)>0$, $\overline{\mu}_1^{E_k}$ is absolutely continuous with respect to $\mu_0$, and consequently, $\supp \overline{\mu}_1^{E_k}\subseteq \supp \mu_0$. Let $K$ denote the indices of the cells $E_k$ for which $F_1(E_k)>0$. Since at most countably many of $E_k$ have positive measure, $K$ is a countable set. Therefore,
\[
\overline{\mu}_1 =\sum_{k\in K}F_1(E_k)\overline{\mu}_1^{E_k}.
\]
Because the support of each $\overline{\mu}_1^{E_k}$ is contained in the support of $\mu_0$ and $\overline{\mu}_1$ is their convex combination, the support of $\overline{\mu}_1$ is contained in the closure of $\bigcup_{k\in K} \supp \overline{\mu}_1^{E_k}$, which is contained in $\supp \mu_0$.\hfill\qed


\subsection*{Proof of Proposition \ref{prop:tail}}
By assumption, there exists a positive $F_1$ measure set of posteriors $H$ such that posteriors in $H$ have uniformly heavier tails than $\mu_0$. Consider an arbitrary measurable partition of the set of posteriors $\Delta(X)$ into sets $\{E_k\}_{k}$. Since $F_1(H)>0$, there exists a cell $E_{k^*}$ of the partition such that $F_1(H\cap E_{k^*})>0$. Since the distributions in $H$ have uniformly heavier tails than $\mu_0$, for every $M$, there exists some $R$ such that, for all $r>R$,
\[
\frac{\mu_1(\Vert x\Vert >r)}{\mu_0(\Vert x\Vert >r)}>M,
\]
for all $\mu_1\in H$. Therefore, for every $M$, there exists some $R$ such that
\begin{align*}
\frac{\overline{\mu}_1^{E_{k^*}}(\Vert x\Vert >r)}{\mu_0(\Vert x\Vert >r)} & = \frac{1}{F_1(E_{k^*})}\int_{E_{k^*}}\frac{\mu_1(\Vert x\Vert >r)}{\mu_0(\Vert x\Vert >r)}F_1(d\mu_1)\\ &  \geq \frac{1}{F_1(E_{k^*})}\int_{H\cap E_{k^*}}\frac{\mu_1(\Vert x\Vert >r)}{\mu_0(\Vert x\Vert >r)}F_1(d\mu_1)\\ & >\frac{MF_1(H\cap E_{k^*})}{F_1(E_{k^*})},
\end{align*}
for any $r>R$. Hence,
\[
\lim_{r\to \infty }\frac{\overline{\mu}_1^{E_{k^*}}(\Vert x\Vert >r)}{\mu_0(\Vert x\Vert >r)} = \infty.
\]
That is, $\overline{\mu}_1^{E_{k^*}}$ has heavier tails than $\mu_0$. Towards a contradiction, suppose $\mu_0$ contains a grain of $\overline{\mu}_1^{E_{k^*}}$. Proposition \ref{prop:unif_abs_cont} then implies that there exists a constant $c\geq 1$ such that $\overline{\mu}_1^{E_{k^*}}(E) \leq c\mu_0(E)$ for any measurable set $E$, a contradiction to the fact that $\overline{\mu}_1^{E_{k^*}}(\Vert x\Vert >r)/\mu_0(\Vert x\Vert >r)$ grows without bound as $r$ goes to infinity. Thus, $\mu_0$ does not contain a grain of $\overline{\mu}_1^{E_{k^*}}$. Since the partition was arbitrary and $F_1(E_{k^*})>0$, by Theorem \ref{thm:main}, the pair $(\mu_0,F_1)$ is not consistent with misspecified Bayesianism.\hfill\qed


\subsection*{Proof of Theorem \ref{thm:normal}}
I first state and prove a lemma.
\begin{lemma}\label{lem:normal_LR}
Let $\mu_0=\mathcal N(m_0,\Sigma_0)$ be an $n$-dimensional multivariate normal distribution with mean $m_0$ and positive definite covariance matrix $\Sigma_0$, and let $E=\{\mathcal N(m_1,\Sigma_1): m_1\in B\}$ be a set of $n$-dimensional multivariate normal distributions with means belonging to the non-empty bounded Borel set $B$ and common positive definite covariance matrix $\Sigma_1$.
\begin{enumerate}[(i)]
    \item Suppose $\Sigma_1\preceq\Sigma_0$ and $m_1-m_0$ is in the image of $\Sigma_0 - \Sigma_1$ for every $m_1\in B$. Then there exists a constant $C$ such that $d\mu_1/d\mu_0\leq C$ for all $\mu_1\in E$.
    \item Suppose $\Sigma_1\not\preceq \Sigma_0$. Then there exist a vector $v\in\mathbb R^n$ and constants $c_1,c_2>0$ such that 
    \[
    \frac{d\mu_1}{d\mu_0}(m_0+tv)\ge c_1 e^{c_2t^2}
    \]
    for all $\mu_1\in E$ and all $t\geq 0$.

    \item Suppose $\Sigma_1\preceq \Sigma_0$ and that there exists a vector $u$ in the kernel of $\Sigma_0-\Sigma_1$ and a constant $\varepsilon>0$ such that $u'(m_1-m_0)\ge \varepsilon$ for all $m_1\in B$. Then there exist a vector $v\in\mathbb R^n$ and constants $c_1,c_2>0$ such that 
    \[
    \frac{d\mu_1}{d\mu_0}(m_0+tv)\ge c_1 e^{c_2 t}
    \]
    for all $\mu_1\in E$ and all $t\geq 0$.
\end{enumerate}
\end{lemma}

\begin{proof}
Since $\mu_0$ and $\mu_1$ have densities with respect to the Lebesgue measure for any $\mu_1\in E$, the Radon--Nikodym derivative $d\mu_1/d\mu_0$ is equal to the ratio of densities. In particular,
\begin{equation}\label{eq:normal_LR}
\frac{d\mu_1}{d\mu_0}(x) =\left(\frac{\det(\Sigma_0)}{\det(\Sigma_1)}\right)^{1/2}\exp\left(-\frac12 (x-m_1)' \Sigma_1^{-1}(x-m_1)+\frac12 (x-m_0)' \Sigma_0^{-1}(x-m_0)\right).  
\end{equation}

To prove part (i), note that this expression is bounded if
\begin{equation}\label{eq:normal_quadratic}
-\frac12 (x-m_1)' \Sigma_1^{-1}(x-m_1)+\frac12 (x-m_0)' \Sigma_0^{-1}(x-m_0)
\end{equation}
is bounded above. This is equivalent to $\Sigma_1^{-1}\succeq \Sigma_0^{-1}$ and $v'(\Sigma_1^{-1}m_1-\Sigma_0^{-1}m_0)=0$ for every vector $v$ such that $(\Sigma_1^{-1}-\Sigma_0^{-1})v=0$. That $\Sigma_1^{-1}\succeq\Sigma_0^{-1}$ follows immediately from the assumption that $\Sigma_0\succeq\Sigma_1$. Next, choose a $v$ such that $(\Sigma_1^{-1}-\Sigma_0^{-1})v=0$. Since
\[
\Sigma_1^{-1}-\Sigma_0^{-1}
=
\Sigma_0^{-1}(\Sigma_0-\Sigma_1)\Sigma_1^{-1},
\]
$\Sigma_1^{-1}v$ is in the kernel of $\Sigma_0-\Sigma_1$. Therefore, since $\Sigma_0-\Sigma_1$ is symmetric, $\Sigma_1^{-1}v$ is orthogonal to any vector in the image of $\Sigma_0-\Sigma_1$. In particular, 
\[
v'(\Sigma_1^{-1}m_1-\Sigma_0^{-1}m_0)=v'\Sigma_1^{-1}(m_1-m_0) = 0,
\]
where the first equality follows from the assumption that $(\Sigma_1^{-1}-\Sigma_0^{-1})v=0$, and the second equality uses the assumption that $m_1-m_0$ is in the image of $\Sigma_0 - \Sigma_1$. Therefore, \eqref{eq:normal_quadratic} is bounded above. In particular,
\begin{align*}
& -\frac12 (x-m_1)' \Sigma_1^{-1}(x-m_1)+\frac12 (x-m_0)' \Sigma_0^{-1}(x-m_0)\\ & \qquad  \leq \frac12 \left(m_1'\Sigma_1^{-1}-m_0'\Sigma_0^{-1}\right)\left(\Sigma_1^{-1}-\Sigma_0^{-1}\right)^\dagger\left(\Sigma_1^{-1}m_1-\Sigma_0^{-1}m_0\right) - \frac12 m_1'\Sigma_1^{-1}m_1 + \frac12 m_0'\Sigma_0^{-1}m_0\\ & \qquad = 
\frac12 (m_1-m_0)' (\Sigma_0-\Sigma_1)^\dagger (m_1-m_0),
\end{align*}
and so, 
\[
\frac{d\mu_1}{d\mu_0}(x) \leq \left(\frac{\det(\Sigma_0)}{\det(\Sigma_1)}\right)^{1/2}\exp\left(
\frac12 (m_1-m_0)' (\Sigma_0-\Sigma_1)^\dagger (m_1-m_0)\right).
\]
Since $(\Sigma_0-\Sigma_1)^\dagger$ is a fixed positive semidefinite matrix, $m_0$ is a fixed vector, and $m_1$ belongs to a bounded set for $\mu_1\in E$, this expression is bounded uniformly for all $\mu_1\in E$.

To prove part (ii), suppose $\Sigma_1\not\preceq \Sigma_0$. Then there exists a vector $v\in\mathbb R^n$ such that
\[
v'\Sigma_0^{-1}v-v'\Sigma_1^{-1}v>0.
\]
Evaluating the likelihood ratio at $x=m_0+tv$ gives
\begin{align}
\log\frac{d\mu_1}{d\mu_0}(m_0+tv)
=&\;
\frac12\log\frac{\det(\Sigma_0)}{\det(\Sigma_1)}
-\frac12 (tv-(m_1-m_0))'\Sigma_1^{-1}(tv-(m_1-m_0))
+\frac12 t^2 v'\Sigma_0^{-1}v \nonumber\\
=&\;
\frac12\log\frac{\det(\Sigma_0)}{\det(\Sigma_1)}
-\frac12 (m_1-m_0)'\Sigma_1^{-1}(m_1-m_0)
+t\,v'\Sigma_1^{-1}(m_1-m_0) \nonumber\\
&
+\frac12 t^2\left(v'\Sigma_0^{-1}v-v'\Sigma_1^{-1}v\right).\label{eq:normal_quadratic_ray}
\end{align}
Since $B$ is bounded, the first two terms on the right-hand side are bounded below uniformly over $m_1\in B$, and the linear term is bounded below by $-Ct$ for some constant $C>0$. Since the coefficient on $t^2$ is strictly positive, there exist constants $c_1,c_2>0$ such that
\[
\frac{d\mu_1}{d\mu_0}(m_0+tv)\ge c_1 e^{c_2 t^2}
\]
for all $\mu_1\in E$ and all $t\geq 0$.

To prove part (iii), assume $\Sigma_1\preceq \Sigma_0$, and suppose there exist a vector $u\in\ker(\Sigma_0-\Sigma_1)$ and a constant $\varepsilon>0$ such that
\[
u'(m_1-m_0)\ge \varepsilon
\]
for all $m_1\in B$. I define
\[
v\equiv \Sigma_1 u,
\]
and evaluate the likelihood ratio at $x=m_0+tv$. Since $(\Sigma_0-\Sigma_1)u=0$, we have $\Sigma_0u=\Sigma_1u=v$, and therefore
\[
v'\Sigma_0^{-1}v=v'u=u'\Sigma_1u=v'\Sigma_1^{-1}v.
\]
Thus, the quadratic term in equation \eqref{eq:normal_quadratic_ray} vanishes. Moreover, $v'\Sigma_1^{-1}(m_1-m_0)=u'(m_1-m_0)$. Hence,
\begin{align*}
\log\frac{d\mu_1}{d\mu_0}(m_0+tv)
&=
\frac12\log\frac{\det(\Sigma_0)}{\det(\Sigma_1)}
-\frac12 (m_1-m_0)'\Sigma_1^{-1}(m_1-m_0)
+t\,u'(m_1-m_0).
\end{align*}
Since $B$ is bounded, the first two terms on the right-hand side are bounded below uniformly over $m_1\in B$. Since $u'(m_1-m_0)\ge \varepsilon$ uniformly over $m_1\in B$, there exist constants $c_1,c_2>0$ such that
\[
\frac{d\mu_1}{d\mu_0}(m_0+tv)\ge c_1 e^{c_2 t}
\]
for all $\mu_1\in E$ and all $t\geq 0$.
\end{proof}


\vspace{1em}

\noindent\emph{Proof of the ``if'' direction.} Partition the image of $\Sigma_0-\Sigma_1$ into a countable collection of bounded Borel sets $\{B_k\}_{k\in\mathbb N}$, and define
\[
E_k\equiv \{\mathcal N(m_1,\Sigma_1): m_1-m_0\in B_k\}.
\]
Because $m_1-m_0$ is in the image of $\Sigma_0-\Sigma_1$ for $F_1$-almost all posteriors, the sets $\{E_k\}_{k\in\mathbb N}$ form a countable measurable partition of $\Delta(X)$ up to a $F_1$-null set of posteriors. Fix a $k$ with $F_1(E_k)>0$. Since $B_k$ is bounded, by part (i) of Lemma \ref{lem:normal_LR}, the Radon--Nikodym derivative $d\mu_1/d\mu_0$ has a uniform bound for all $\mu_1$ in $E_k$. Therefore, $d\overline{\mu}_1^{E_k}/d\mu_0$ is bounded. Proposition \ref{prop:unif_abs_cont} then implies that $\mu_0$ contains a grain of $\overline{\mu}_1^{E_k}$. Since this holds for any $k$ with $F_1(E_k)>0$, Theorem \ref{thm:main} implies that $(\mu_0,F_1)$ is consistent with misspecified Bayesianism.

\vspace{1em}

\noindent\emph{Proof of the ``only if'' direction.} Suppose first that $\Sigma_1\not\preceq \Sigma_0$. Let $\{E_k\}_{k\in\mathbb N}$ be an arbitrary countable measurable partition of $\Delta(X)$, and fix some $k$ such that $F_1(E_k)>0$. There exists a non-empty bounded Borel set $B\subseteq \mathbb R^n$ such that the set
\[
\hat E_k\equiv \{\mathcal N(m_1,\Sigma_1)\in E_k:m_1\in B\}
\]
is a measurable subset of $E_k$ with positive $F_1$ measure. By part (ii) of Lemma \ref{lem:normal_LR}, there exist a vector $v\in\mathbb R^n$ and constants $c_1,c_2>0$ and $T<\infty$ such that
\[
\frac{d\mu_1}{d\mu_0}(m_0+tv)\ge c_1 e^{c_2t^2}
\]
for all $t\ge T$ and all $\mu_1\in \hat E_k$. Therefore,
\begin{align*}
\frac{d\overline\mu_1^{E_k}}{d\mu_0}(m_0+tv)
&=
\frac{1}{F_1(E_k)}\int_{E_k}\frac{d\mu_1}{d\mu_0}(m_0+tv)\,F_1(d\mu_1)\\
&\ge
\frac{1}{F_1(E_k)}\int_{\hat E_k}\frac{d\mu_1}{d\mu_0}(m_0+tv)\,F_1(d\mu_1)\\
&\ge
\frac{F_1(\hat E_k)}{F_1(E_k)}c_1 e^{c_2t^2}.
\end{align*}
Hence, $\frac{d\overline\mu_1^{E_k}}{d\mu_0}(x)$ goes to infinity along the ray $m_0 + tv$. The facts that $\frac{d\overline\mu_1^{E_k}}{d\mu_0}(x)$ is a continuous function of $x$ and $\mu_0$ has a strictly positive density on $\mathbb R^n$ then imply that there exists no $C$ such that $\frac{d\overline\mu_1^{E_k}}{d\mu_0}(x)\leq C$ for $\mu_0$-almost all $x$. Therefore, Proposition \ref{prop:unif_abs_cont} implies that $\mu_0$ does not contain a grain of $\overline\mu_1^{E_k}$. Since the partition was arbitrary, Theorem \ref{thm:main} implies that $(\mu_0,F_1)$ is not consistent with misspecified Bayesianism.

Next suppose $\Sigma_1\preceq \Sigma_0$, but $m_1-m_0$ is not in the image of $\Sigma_0-\Sigma_1$ for a set of posteriors that are realized with positive $F_1$ probability. Since $\Sigma_0-\Sigma_1$ is symmetric, there exists a vector $u\in\ker(\Sigma_0-\Sigma_1)$ such that
\[
u'(m_1-m_0)\neq 0
\]
for a set of posteriors $\mu_1=\mathcal{N}(m_1,\Sigma_1)$ with positive $F_1$ probability. Hence, at least one of the sets $\left\{\mathcal{N}(m_1,\Sigma_1):u'(m_1-m_0)>0\right\}$ and $\left\{\mathcal{N}(m_1,\Sigma_1):u'(m_1-m_0)<0\right\}$ has positive $F_1$ probability. Without loss of generality, suppose the first one does. Then
\[
\{\mathcal{N}(m_1,\Sigma_1):u'(m_1-m_0)>0\}
=
\bigcup_{j=1}^\infty \{\mathcal{N}(m_1,\Sigma_1):u'(m_1-m_0)\ge 1/j\},
\]
so there exists $\varepsilon>0$ such that
\[
\{\mathcal{N}(m_1,\Sigma_1):u'(m_1-m_0)\ge \varepsilon\}
\]
is a measurable set with positive $F_1$ measure. Let $\{E_k\}_{k\in\mathbb N}$ be an arbitrary countable measurable partition of $\Delta(X)$. Since the above set has positive probability, there exists some $k$ such that
\[
F_1\big(E_k\cap \{\mathcal{N}(m_1,\Sigma_1):u'(m_1-m_0)\ge \varepsilon\}\big)>0.
\]
Fix such a $k$. As above, there exists a non-empty bounded Borel set $B\subseteq\mathbb R^n$ such that the set 
\[
\hat E_k\equiv E_k\cap \{\mathcal N(m_1,\Sigma_1): u'(m_1-m_0)\ge \varepsilon,\ m_1\in B\}
\]
is a measurable set with positive $F_1$ probability. By part (iii) of Lemma \ref{lem:normal_LR}, there exist a vector $v\in\mathbb R^n$ and constants $c_1,c_2>0$ and $T<\infty$ such that
\[
\frac{d\mu_1}{d\mu_0}(m_0+tv)\ge c_1 e^{c_2t}
\]
for all $t\ge T$ and all $\mu_1\in \hat E_k$. Therefore,
\begin{align*}
\frac{d\overline\mu_1^{E_k}}{d\mu_0}(m_0+tv)
&=
\frac{1}{F_1(E_k)}\int_{E_k}\frac{d\mu_1}{d\mu_0}(m_0+tv)\,F_1(d\mu_1)\\
&\ge
\frac{1}{F_1(E_k)}\int_{\hat E_k}\frac{d\mu_1}{d\mu_0}(m_0+tv)\,F_1(d\mu_1)\\
&\ge
\frac{F_1(\hat E_k)}{F_1(E_k)}c_1 e^{c_2t}.
\end{align*}
Hence, $\frac{d\overline\mu_1^{E_k}}{d\mu_0}(x)$ goes to infinity along the ray $m_0 + tv$. The facts that $\frac{d\overline\mu_1^{E_k}}{d\mu_0}(x)$ is a continuous function of $x$ and $\mu_0$ has a strictly positive density on $\mathbb R^n$ then imply that there exists no $C$ such that $\frac{d\overline\mu_1^{E_k}}{d\mu_0}(x)\leq C$ for $\mu_0$-almost all $x$, so by Proposition \ref{prop:unif_abs_cont}, $\mu_0$ does not contain a grain of $\overline\mu_1^{E_k}$. Since this happens for some positive-probability cell of every countable measurable partition, the condition in Theorem \ref{thm:main} fails. Therefore, $(\mu_0,F_1)$ is not consistent with misspecified Bayesianism.\hfill\qed


\subsection*{Proof of Theorem \ref{thm:main-x}}
The proof closely follows the proof of Theorem \ref{thm:main}.
\vspace{1em}

\noindent\emph{Proof of the ``if'' direction.} Given the true distribution $\mathbb{P}\in\Delta(X\times S)$, I construct the subjective distribution $\mathbb{Q}\in\Delta(X\times S)$ that rationalizes an observed pair $\big(\mu_0,\{F_{1x}\}_{x\in X}\big)$ satisfying the assumption of the theorem. By assumption, $F_{1x} = \mathbb{P}(\cdot|x) \circ \varphi^{-1}$ for some mapping $\varphi$ and for all $x\in X$. Therefore, for any $E\in \mathcal{S}$,
\[
\overline{F}_1(E) = \int_{X}F_{1x}(E)\mathbb{P}_X(dx) = \int_X \mathbb{P}\left(\varphi^{-1}(E)\big|x\right)\mathbb{P}_X(dx)=\mathbb{P}_S\left(\varphi^{-1}(E)\right),
\]
and so $\overline{F}_1=\mathbb{P}_S\circ \varphi^{-1}$. Furthermore, by assumption, there exists a countable measurable partition $\{E_k\}_{k}$ of the set of posteriors $\Delta(X)$ into sets such that, for every $E_k$ with $\overline{F}_1(E_k)>0$, $\mu_0$ contains a grain of the conditional average posterior $\overline{\overline{\mu}}_1^{E_k}$ given $E_k$. Let $K$ denote the indices of the cells $E_k$ for which $\overline{F}_1(E_k)>0$. For any $k\in K$, since $\mu_0$ contains a grain of $\overline{\overline{\mu}}_1^{E_k}$, there exist some $\epsilon_k\in(0,1]$ and some probability measure $\mu'_k\in \Delta(X)$ such that $\mu_0=\epsilon_k\overline{\overline{\mu}}_1^{E_k}+(1-\epsilon_k)\mu'_k$. Define $\hat{S}\equiv \bigcup_{k\in K} E_k$, and note that $\overline{F}_1(\hat{S})=1$.

I construct $\mathbb{Q}$ as in the proof of the ``if'' direction of Theorem \ref{thm:main} with $F_1$ replaced by $\overline{F}_1$ throughout. In particular, I let $\ominus\in S$ denote a signal such that $\mathbb{P}_S(\{\ominus\})=0$; for any $s\in S$ such that $\varphi(s)\in\hat{S}$ and $s\neq \ominus$, I set $\mathbb{Q}(D|s)=\varphi(s)(D)$ for all $D\in\mathcal{X}$; I set $\mathbb{Q}(\cdot|\ominus)=\mu'$, where $\mu'$ is an arbitrary probability measure over $X$ if $1-\sum_{k\in K}\epsilon_k\overline{F}_1(E_k)=0$ and is given by 
\[
\mu'(D)\equiv \sum_{k\in K}\frac{(1-\epsilon_k)\overline{F}_1(E_k)}{1-\sum_{k\in K}\epsilon_k\overline{F}_1(E_k)}\mu'_k(D)
\]
for all $D\in\mathcal{X}$ if $1-\sum_{k\in K}\epsilon_k\overline{F}_1(E_k)>0$; I set $\mathbb{Q}(D|s)=\mu_0(D)$ for any $s\in S$ such that $s\neq \ominus$ and $\varphi(s)\notin \hat{S}$ and all $D\in\mathcal{X}$; and I let 
\begin{equation}
\mathbb{Q}_S(E)\equiv\sum_{k\in K}\epsilon_k\mathbb{P}_S\left(E\cap \varphi^{-1}(E_k)\right)+\left(1-\sum_{k\in K}\epsilon_k\overline{F}_1(E_k)\right)\mathbb{1}\{\ominus\in E\},\label{eq:Q_S_definition-x}
\end{equation}
for all $E\in\mathcal{S}$ and let
\begin{equation}\label{eq:Bayes_rule_proof-x}
\mathbb{Q}(D\times E) \equiv \int_E \mathbb{Q}(D|s)\mathbb{Q}_S(ds)
\end{equation}
for all $D\in\mathcal{X}$ and $E\in \mathcal{S}$.

By an identical argument to that in the proof of Theorem \ref{thm:main}, $\mathbb{P}_S$ is absolutely continuous with respect to $\mathbb{Q}_S$ and $\mathbb{Q}_X=\mu_0$. I only need to show that $\mathbb{P}(\cdot|x)\circ\varphi_{\mathbb{Q}}^{-1}=F_{1x}$ for $\mathbb{P}_X$-almost all $x$. By definition,  $F_{1x}=\mathbb{P}(\cdot|x)\circ\varphi^{-1}$ for all $x$. On the other hand, since $\mathbb{P}_S(\varphi^{-1}(\hat{S})\setminus\{\ominus\})=1$, for $\mathbb{P}_X$-almost all $x$,
\[
\mathbb{P}(\varphi^{-1}(\hat{S})\setminus\{\ominus\}|x)=1.
\]
Therefore, for any $E\in\mathcal{S}$ and $\mathbb{P}_X$-almost all $x$,
\begin{align*}
\left(\mathbb{P}(\cdot|x)\circ \varphi_{\mathbb{Q}}^{-1}\right)(E) & =\mathbb{P}\left(\{s\in S:\mathbb{Q}(\cdot|s)\in E\}|x\right)\\ & =\mathbb{P}\left(\{s\in S:\mathbb{Q}(\cdot|s)\in E,\varphi(s)\in\hat{S},s\neq \ominus\}|x\right)\\ & =\mathbb{P}\left(\{s\in S:\varphi(s)\in E\}|x\right)=F_{1x}(E).
\end{align*}
This completes the proof of the first direction.

\vspace{1em}
\noindent\emph{Proof of the ``only if'' direction.} Let $\mathbb{Q}$ denote the subjective distribution on $X\times S$, and let $\mathbb{Q}(\cdot|\cdot)$ denote a regular conditional probability of $\mathbb{Q}$ given $\mathcal{S}$. Since the signal labels have no inherent meaning, I assume without loss of generality that $\mathbb{Q}(D|\mu)=\mu(D)$ for any $\mu\in S=\Delta(X)$. Given those signal labels, $\varphi_{\mathbb{Q}}$ is the identity mapping, and $F_{1x}=\mathbb{P}(\cdot|x)\circ \varphi_{\mathbb{Q}}^{-1}=\mathbb{P}(\cdot|x)$. Therefore, 
\[
\overline{F}_{1}\equiv \int_X F_{1x}\mathbb{P}_X(dx) = \int_X \mathbb{P}(\cdot|x)\mathbb{P}_X(dx) = \mathbb{P}_S.
\]

Since $\mathbb{Q}$ satisfies condition (a) of Definition \ref{def:Bayesianism-x} and $\mathbb{Q}(\cdot|\cdot)$ is a regular conditional probability of $\mathbb{Q}$ given $\mathcal{S}$,
\begin{equation}
\mu_0(D) = \mathbb{Q}_X(D) = \int_{S}\mathbb{Q}(D|s)\mathbb{Q}_S(ds)\label{eq:onlyif1-x}
\end{equation}
for all $D\in\mathcal{X}$. On the other hand, by condition (b) of Definition \ref{def:Bayesianism-x}, $\mathbb{P}_S$ is absolutely continuous with respect to $\mathbb{Q}_S$. Hence, by the Radon--Nikodym theorem, there exists a Radon--Nikodym derivative $f\equiv \frac{d\mathbb{P}_S}{d\mathbb{Q}_S}$. For $k\in \mathbb{N}$, define
\[
E_k \equiv \big\{s\in S:f(s)\in[k-1,k)\big\}.
\]
Since $f$ is a measurable function, $E_k$ is a measurable subset of $S$ for any $k\in\mathbb{N}$. Furthermore, the sets $\{E_k\}_{k\in\mathbb N}$ form a countable measurable partition of $S=\Delta(X)$. For any $k$ such that $\overline{F}_1(E_k)>0$,
\begin{equation*}
\overline{\overline{\mu}}_1^{E_k}=\frac{1}{\overline{F}_1(E_k)}\int_{E_k} \mu \overline{F}_1(d\mu)=\frac{1}{\overline{F}_1(E_k)}\int_{E_k} \mu \;\mathbb{P}_S\circ\varphi_{\mathbb{Q}}^{-1}(d\mu)=\frac{1}{\overline{F}_1(E_k)}\int_{\varphi_{\mathbb{Q}}^{-1}(E_k)}\mathbb{Q}(\cdot|s)\mathbb{P}_S(ds).
\end{equation*}
Therefore, since $\varphi_{\mathbb{Q}}^{-1}(E_k)=E_k$,
\begin{equation}
\overline{\overline{\mu}}_1^{E_k}(D) = \frac{1}{\overline{F}_1(E_k)}\int_{E_k}\mathbb{Q}(D|s)\mathbb{P}_S(ds)\label{eq:onlyif2-x}    
\end{equation}
for any $D\in\mathcal{X}$.
Since $f$ is the Radon--Nikodym derivative of $\mathbb{P}_S$ with respect to $\mathbb{Q}_S$,
\begin{equation}
\int_{E_k}\mathbb{Q}(D|s)\mathbb{P}_S(ds) = \int_{E_k}\mathbb{Q}(D|s)f(s)\mathbb{Q}_S(ds) \leq k \int_{E_k}\mathbb{Q}(D|s)\mathbb{Q}_S(ds)\leq k \int_{S}\mathbb{Q}(D|s)\mathbb{Q}_S(ds),\label{eq:onlyif3-x}    
\end{equation}
where the first inequality is by the definition of set $E_k$, and the second inequality is due to the fact that $\int_{S\setminus E_k}\mathbb{Q}(D|s)\mathbb{Q}_S(ds)\geq 0$. Equations \eqref{eq:onlyif1-x}--\eqref{eq:onlyif3-x} imply
\[
\overline{\overline{\mu}}_1^{E_k}(D) = \frac{1}{\overline{F}_1(E_k)}\int_{E_k}\mathbb{Q}(D|s)\mathbb{P}_S(ds)\leq \frac{k}{\overline{F}_1(E_k)}\int_{S}\mathbb{Q}(D|s)\mathbb{Q}_S(ds) = \frac{k}{\overline{F}_1(E_k)}\mu_0(D).
\]
The $k/\overline{F}_1(E_k)$ constant in the above inequality is independent of $D$. Therefore, by Proposition~\ref{prop:unif_abs_cont}, $\mu_0$ contains a grain of $\overline{\overline{\mu}}_1^{E_k}$.\hfill\qed


\subsection*{Proof of Theorem \ref{thm:Kamenica_Gentzkow} }

\noindent\emph{Proof of the ``if'' direction.} The proof of this direction is constructive. I define the regular conditional probability $\mathbb{Q}(\cdot|\cdot):\mathcal{X}\times S\to [0,1]$ by setting $\mathbb{Q}(D|s)=\varphi(s)(D)$ for all $s\in S$ and all $D\in\mathcal{X}$. By construction, $\mathbb{Q}(\cdot|s)$ is a probability distribution on $(X,\mathcal{X})$ for any $s$, and the mapping $s\mapsto \mathbb{Q}(D|s)$ is measurable for all $D\in\mathcal{X}$. I set the $S$-marginal $\mathbb{Q}_S$ of the subjective distribution equal to the true distribution $\mathbb{P}_S$ of signals and define $\mathbb{Q}$ as in \eqref{eq:Bayes_rule_proof} with $\mathbb{Q}_S=\mathbb{P}_S$. By construction, $\mathbb{Q}(\cdot|\cdot)$ is a regular conditional probability of $\mathbb{Q}$ given $\mathcal{S}$. Next, note that for any $D\in\mathcal{X}$,
\[
\mathbb{Q}_X(D) = \int_S \mathbb{Q}(D|s)\mathbb{P}_S(ds) = \int_S \mu(D)F_1(d\mu)=\overline{\mu}_1(D)=\mu_0(D),
\]
where the last equality follows the assumption that $\overline{\mu}_1=\mu_0$. Moreover, by an argument similar to the one in the proof of Theorem \ref{thm:main}, 
\[
\mathbb{P}_S\circ \varphi_{\mathbb{Q}}^{-1}(E) = \mathbb{P}_S\left(\{s\in S:\mathbb{Q}(\cdot|s)\in E\}\right) =\mathbb{P}_S\left(\{s\in S:\varphi(s)\in E\}\right)=F_1(E)
\]
for all $E\in\mathcal{S}$. This shows that the subjective distribution $\mathbb{Q}$ constructed above rationalizes the observed pair $(\mu_0,F_1)$.

\vspace{1em}
\noindent\emph{Proof of the ``only if'' direction.}
By the argument in the proof of the ``only if'' direction of Theorem \ref{thm:main}, $\mu_0(D) =\int_{S}\mathbb{Q}(D|s)\mathbb{Q}_S(ds)$ and $\overline{\mu}_1(D) =\int_S\mathbb{Q}(D|s)\mathbb{P}_S(ds)$ for any $D\in\mathcal{X}$. The assumption that $\mathbb{Q}_S=\mathbb{P}_S$ completes the proof.\hfill\qed


\subsection*{Proof of Theorem \ref{thm:Shmaya_Yariv}}
\noindent\emph{Proof of (i) $\implies$ (ii).} Suppose $(\mu_0,F_1)$ is consistent with Bayesianism given a subjective distribution $\mathbb{Q}$ with an $S$-marginal $\mathbb{Q}_S$ that is equivalent to $\mathbb{P}_S$, and let $\mathbb{Q}(\cdot|\cdot)$ denote the regular conditional probability of $\mathbb{Q}$ given $\mathcal{S}$. I define $\lambda\in\Delta(S)=\Delta(\Delta(X))$ as follows:
\[
\lambda(E) \equiv  \mathbb{Q}_S(\{s\in S: \mathbb{Q}(\cdot|s)\in E\})
\]
for all $E\in\mathcal{S}$. I next show that $\lambda$ and $F_1$ are equivalent and $\mu_0=\int \mu\lambda(d\mu)$. Since $\mathbb{Q}$ satisfies condition (a) and $\mathbb{Q}(\cdot|\cdot)$ is a regular conditional probability of $\mathbb{Q}$ given $\mathcal{S}$,
\[
\mu_0 = \mathbb{Q}_X = \int_{S}\mathbb{Q}(\cdot|s)\mathbb{Q}_S(ds)=\int_S \mu\lambda(d\mu),
\]
where the last equality is by the definition of $\lambda$. On the other hand, for all $E\in\mathcal{S}$,
\[
\lambda(E)=\mathbb{Q}_S(\{s\in S: \mathbb{Q}(\cdot|s)\in E\})
\]
and 
\[
F_1(E) = \mathbb{P}_S\circ \varphi_{\mathbb{Q}}^{-1}(E) = \mathbb{P}_S(\{s\in S: \mathbb{Q}(\cdot|s)\in E\}),
\]
where the first equality above is by condition (c) of Definition \ref{def:Bayesianism}. Since $\mathbb{Q}_S$ and $\mathbb{P}_S$ are equivalent, so are $\lambda$ and $F_1$.

\vspace{1em}
\noindent\emph{Proof of (ii) $\implies$ (i).}
Suppose there exists a probability measure $\lambda \in \Delta(\Delta(X))$ such that $\lambda$ and $F_1$ are equivalent and $\mu_0=\int \mu\lambda(d\mu)$. By the Radon--Nikodym theorem, there are derivatives $f\equiv \frac{d\lambda}{dF_1}$ and $\frac{1}{f}\equiv \frac{dF_1}{d\lambda}$. Set $\mathbb{Q}(D|s)=\varphi(s)(D)$ for all $s\in S$ and $D\in\mathcal{X}$, and set $\mathbb{Q}_S(ds)=f(\varphi(s))\mathbb{P}_S(ds)$. I need to show that $\mathbb{Q}_S$, as defined above, is indeed a probability distribution on $(S,\mathcal{S})$. By construction, $\mathbb{Q}_S(E)\geq 0$ for all $E\in\mathcal{S}$, and $\mathbb{Q}_S(\emptyset)=0$. Next, note that
\[
\int_S \mathbb{Q}_S(ds) = \int_S f(\varphi(s))\mathbb{P}_S(ds)=\int_S f(\mu) \mathbb{P}_S\circ \varphi^{-1}(d\mu)=\int_Sf(\mu)F_1(d\mu)=\int_S \lambda(d\mu)=1,
\]
where the first equality is by definition, the second one uses the change-of-variables formula for pushforward measures, the third equality is due to the fact that $F_1=\mathbb{P}_S\circ\varphi^{-1}$, the fourth one uses the definition of $f$, and the last equality is because $\lambda$ is a probability measure on $S$. Finally, $\mathbb{Q}_S$ is countably additive since $\mathbb{P}_S$ is countably additive. Therefore, $\mathbb{Q}_S$ is a well-defined probability distribution. I finish the construction by defining $\mathbb{Q}$ as in equation \eqref{eq:Bayes_rule_proof}. Note that by construction, $\mathbb{Q}(\cdot|\cdot)$ is a regular conditional probability of $\mathbb{Q}$ given $\mathcal{S}$. Furthermore, by an argument similar to the one in the above display,
\begin{align*}
\mathbb{Q}_X & =\int_S \mathbb{Q}(\cdot|s)\mathbb{Q}_S(ds)=\int_S\varphi(s)f(\varphi(s))\mathbb{P}_S(ds)= \int_S \mu f(\mu)F_1(d\mu) = \int_S \mu \lambda(d\mu)=\mu_0,
\end{align*}
where the last equality is by assumption. Therefore, condition (a) of Definition \ref{def:Bayesianism} is satisfied. Furthermore, since $\mathbb{Q}_S(ds)=f(\varphi(s))\mathbb{P}_S(ds)$ and $\mathbb{P}_S(ds)=\frac{1}{f(\varphi(s))}\mathbb{Q}_S(ds)$, probability distributions $\mathbb{Q}_S$ and $\mathbb{P}_S$ are equivalent. That is, condition (b) of Definition \ref{def:Bayesianism} is satisfied, and $\mathbb{Q}_S$ is absolutely continuous with respect to $\mathbb{P}_S$. On the other hand,
\[
\mathbb{P}_S\circ \varphi_{\mathbb{Q}}^{-1}(E) = \mathbb{P}_S\left(\{s\in S:\mathbb{Q}(\cdot|s)\in E\}\right) =\mathbb{P}_S\left(\{s\in S:\varphi(s)\in E\}\right)=\mathbb{P}_S\circ\varphi^{-1}(E)=F_1(E)
\]
for all $E\in\mathcal{S}$, implying that condition (c) is also satisfied.\hfill\qed


\section{An example with a finite state space}\label{sec:discrete_example}

This example illustrates the proof of the ``if'' direction of Theorem \ref{thm:main} in a discrete-state setting. The state takes values in the set $X=\{H,L\}$. The prior is the uniform distribution over $X$. The distribution of posteriors $F_1$ is as follows: with a one-quarter probability, the belief that the state is $H$ goes to $0.8$; with the remaining three-quarters probability, the belief that the state is $H$ goes up to $1.0$. Is this belief sequence consistent with misspecified Bayesianism? The answer is yes. This conclusion follows from Corollary \ref{cor:finite_full_support} by noting that $\mu_0$ has full support over~$X$. In what follows, I illustrate how $(\mu_0,F_1)$ can be rationalized.

$F_1$ imposes some restrictions on the true distribution of signals $\mathbb{P}_S$ and the mapping $\varphi$ used to form beliefs. Since the posterior takes on two values, there are at least two signals that are realized with positive probability. The observation of one set of signals moves the belief that the state is $H$ to $0.8$. Since with a one-quarter probability, the posterior is $\mu_1(\{H\})=0.8$, the signals that lead to this posterior must have probability $\mathbb{P}_S(\{s:\varphi(s)=(\mu_1(\{H\})=0.8)\})=0.25$. Likewise, there is a set of signals that has true probability $\mathbb{P}_S(\{s:\varphi(s)=(\mu_1(\{H\})=1)\})=0.75$ and leads to the posterior that the state is $H$ with certainty. With slight abuse of notation, I refer to the $\{s:\varphi(s)=(\mu_1(\{H\})=0.8)\}$ and $\{s:\varphi(s)=(\mu_1(\{H\})=1)\}$ events simply as the $s=0.8$ and $s=1$ signals, respectively.\footnote{This is not an abuse of notation under the assumption that $\varphi$ is the identity mapping. Note that the assumption that $\varphi$ is the identity mapping is innocuous in this example since $F_1$ only identifies $\mathbb{P}_S\circ \varphi^{-1}=F_1$---but not $\mathbb{P}_S$ or 
$\varphi$. Nonetheless, the construction in the example can be easily modified to allow for the possibility that $F_1$ and $\mathbb{P}_S$ are separately identified. I do not pursue this extension here since it would lead to additional notational complexity without offering any new insights. See the proof of Theorem \ref{thm:main} for the general construction.}

I illustrate how $(\mu_0,F_1)$ can be rationalized by finding a subjective distribution $\mathbb{Q}$ such that the belief sequence of a Bayesian agent with subjective distribution $\mathbb{Q}$ matches the prior and the distribution of posteriors. The distribution $\mathbb{Q}$ needs to satisfy three requirements for it to rationalize the prior $\mu_0$ and posterior distribution $F_1$. First, $\mathbb{Q}$ must be consistent with $\mu_0$; i.e., $\mathbb{Q}_X(\{H\})=\mu_0(\{H\})=0.5$.  Second, it must assign positive probability to the $s=0.8$ and $s=1$ signals for Bayes' rule to be applicable after the observation of those signals. Third, the posterior conditional on the $s=0.8$ and $s=1$ signals must be consistent with the corresponding posteriors; i.e., $\mathbb{Q}(\{H\}|s=0.8)=0.8$ and $\mathbb{Q}(\{H\}|s=1.0)=1.0$. 

One also needs to specify the subjective probability of observing signals other than $0.8$ and $1.0$. I start by assuming that, according to $\mathbb{Q}$, the signal can only take values $s=0.8$ and $s=1.0$. This assumption constrains $\mathbb{Q}$ to satisfy $\mathbb{Q}(\{(x,s):s\in \{0.8,1.0\}\})=1$. This constraint, together with the requirements previously discussed and the requirement that $\mathbb{Q}(x,s)\geq 0$ for any $(x,s)$, yields a mixed system of equalities and inequalities for the four unknown probabilities $\mathbb{Q}(H,0.8)$, $\mathbb{Q}(L,0.8)$, $\mathbb{Q}(H,1.0)$, and $\mathbb{Q}(L,1.0)$:
\begin{align}
    & \mathbb{Q}(H,0.8)+\mathbb{Q}(L,0.8)>0,\label{eq:eg_requirement_1}\\
    & \mathbb{Q}(H,1.0)+\mathbb{Q}(L,1.0)>0,\\
    & \frac{\mathbb{Q}(H,0.8)}{\mathbb{Q}(H,0.8)+\mathbb{Q}(L,0.8)}=0.8,\\
    & \frac{\mathbb{Q}(H,1.0)}{\mathbb{Q}(H,1.0)+\mathbb{Q}(L,1.0)}=1.0,\label{eq:eg_requirement_4}\\
    &  \mathbb{Q}(H,0.8), \mathbb{Q}(L,0.8), \mathbb{Q}(H,1.0), \mathbb{Q}(L,1.0) \geq 0,\label{eq:eg_requirement_7}\\
    & \mathbb{Q}(H,0.8)+\mathbb{Q}(H,1.0)=0.5,\label{eq:eg_sum_to_half_H}\\
    & \mathbb{Q}(L,0.8)+\mathbb{Q}(L,1.0)=0.5.\label{eq:eg_sum_to_half_L}
\end{align}
It is easy to verify that this system does not have a solution. 

Thus, for the belief sequence to be consistent with misspecified Bayesianism, the subjective distribution must entertain the possibility that the signal takes values outside the set $\{0.8,1.0\}$. With slight abuse of notation, I let $s=\ominus$ denote the event that the signal takes a value outside the set $\{0.8,1.0\}$. Constraints \eqref{eq:eg_sum_to_half_H} and \eqref{eq:eg_sum_to_half_L} must now be modified as follows:
\begin{align}
    & \mathbb{Q}(H,0.8)+\mathbb{Q}(H,1.0)+\mathbb{Q}(H,\ominus)=0.5,\label{eq:eg_sum_to_half_H_prime}\\
    & \mathbb{Q}(L,0.8)+\mathbb{Q}(L,1.0)+\mathbb{Q}(L,\ominus)=0.5.\label{eq:eg_sum_to_half_L_prime}
\end{align}
The remaining requirements, expressed in equations  \eqref{eq:eg_requirement_1}--\eqref{eq:eg_requirement_7}, remain intact. However, $\mathbb{Q}$ must now additionally satisfy the two non-negativity requirements:
\begin{align}
    & \mathbb{Q}(H,\ominus),\mathbb{Q}(L,\ominus)\geq 0.\label{eq:eg_requirement_5}
\end{align}
Equations \eqref{eq:eg_requirement_1}--\eqref{eq:eg_requirement_7} and \eqref{eq:eg_sum_to_half_H_prime}--\eqref{eq:eg_requirement_5} constitute a mixed system of equalities and inequalities for the six unknown probabilities $\mathbb{Q}(H,0.8)$, $\mathbb{Q}(L,0.8)$, $\mathbb{Q}(H,1.0)$, $\mathbb{Q}(L,1.0)$, $\mathbb{Q}(H,\ominus)$, and $\mathbb{Q}(L,\ominus)$. The fact that the average posterior has the same support as the prior is sufficient to ensure that this system has a solution. One such solution---and the one corresponding to the proof of Theorem \ref{thm:main}---is as follows:
\settowidth{\mycolwd}{$0.1875$}
\[
\renewcommand\arraystretch{2}
\begin{array}{l|C{\mycolwd}|C{\mycolwd}|C{\mycolwd}|}
\multicolumn{1}{r}{}
 & \multicolumn{1}{c}{0.8}
 & \multicolumn{1}{c}{1.0}
 & \multicolumn{1}{c}{\ominus}\\
\cline{2-4}
H & \hfill0.25\hfill\hfill  & \hfill0.25\hfill\hfill & \hfill0\hfill\hfill\\
\cline{2-4}
L & 0.0625 & \hfill 0 \hfill\hfill & \hfill 0.4375\hfill\\
\cline{2-4}
\end{array}
\]

Note that the pair $(\mu_0,F_1)$ can be rationalized only if we allow for a misspecified belief about the distribution of signals. If the subjective distribution were to agree with the true distribution on the probabilities of different signals, the system of equalities and inequalities that determines $\mathbb{Q}$ would have no solution.


\newpage

\fontsize{11}{11}\selectfont\baselineskip0.52cm
\bibliographystyle{ecta-fullname}
\bibliography{subjective}

\end{document}